\documentclass[twocolumn,showpacs,showkeys]{revtex4}
\usepackage{graphicx,subfigure}
\input{psfig.sty}

\newcommand{\bastar}{\begin{eqnarray*}}
\newcommand{\eastar}{\end{eqnarray*}}
\newskip\humongous \humongous=0pt plus 1000pt minus 1000pt

\newif\ifdtup

\relax
\newcommand{\be}{\begin{equation}}
\newcommand{\ee}{\end{equation}}
\newcommand{\bea}{\begin{eqnarray}}
\newcommand{\eea}{\end{eqnarray}}
\newcommand{\X}{{\vec X}}
\newcommand{\pro}{\partial}
\newcommand{\n}{\hat n}
\newcommand{\oneg}{\displaystyle\frac{1}{g}}

\newcommand{\D}{{\hat D}}

\newcommand{\A}{{\vec A}}
\newcommand{\valpha}{{\vec \alpha}}

\newcommand{\dfrac}{\displaystyle\frac}
\newcommand{\ba}{\begin{array}}
\newcommand{\ea}{\end{array}}

\newcommand{\nn}{\nonumber}
\newcommand{\hn}{\hat n}
\begin{document}
\title{Topological Objects in Two-gap Superconductor:I}
\bigskip
\author{Y. M. Cho}
\email{ymcho@yongmin.snu.ac.kr}
\author{Pengming Zhang}
\date{\today}
\email{zhpm@phya.snu.ac.kr}
\affiliation{Center for Theoretical Physics and School of Physics \\
College of Natural Sciences, Seoul National University,
Seoul 151-742, Korea  \\}
\begin{abstract}
We discuss topological objects, in particular the non-Abrikosov
vortex and the magnetic knot made of the
twisted non-Abrikosov vortex, in two-gap superconductor.
We show that there are two types of non-Abrikosov vortex in
Ginzburg-Landau theory of two-gap superconductor, the D-type which has
no concentration of the condensate at the core
and the N-type which has a non-trivial
profile of the condensate at the core, under a wide
class of realistic interaction potential.
Furthermore, we show that we can construct a stable magnetic knot
by twisting the non-Abrikosov vortex and connecting two periodic ends
together, whose knot topology $\pi_3(S^2)$ is described by the Chern-Simon
index of the electromagnetic potential.
We discuss how these topological objects can be constructed in
$\rm MgB_2$ or in liquid metallic hydrogen.
\end{abstract}
\maketitle

\section{Introduction}

Topological objects, in particular finite energy topological
objects (monopoles, vortices, skyrmions, and knots),
have played increasingly important role
in physics \cite{dirac,abri,skyr,fadd1,prl01}. In condensed matter the
best known topological objects are the Abrikosov
vortex in one-gap superconductors and similar ones in Bose-Einstein
condensates and superfluids, which have been the subject
of intensive studies.
A recent advent of two-component Bose-Einstein condensates
and two-gap superconductors \cite{bec,sc},
however, has opened up an exciting new possibility
for us to construct far more interesting topological
objects in laboratories.
It has already been shown that non-Abrikosov vortices whose
topology is fixed by $\pi_2(S^2)$ and finite energy topological
knots whose topology is fixed by $\pi_3(S^2)$ exist in these
condensed matters \cite{ijpap,ruo,pra05,prb05,baba1,cm2}.
The reason for this is
that these condensed matters are made of
two components which can be viewed as an $SU(2)$ multiplet.
In general this type of topological objects is possible
when one has a multi-component condensates,
which allows the non-Abelian topology.

The purpose of this paper is to discuss new
topological objects in Ginzburg-Landau theory of
two-gap superconductor in detail.
{\it With a most general $U(1)\times U(1)$ symmetric potential
which can describe a wide class of
two-gap superconductors we first show that there are
two types of non-Abrikosov vortex,
D-type and N-type, in two-gap superconductor.
The D-type has no concentration of the condensate
at the core, but the N-type has a non-trivial
profile of the condensate at the core.  The reason why the
two-gap superconductor has two types of vortex is
that the vortex in two-gap superconductor allows
two different boundary conditions. In terms of topology
the non-Abrikosov vortex is described by two
types of topology, non-Abelian $\pi_2(S^2)$ topology
or Abelian $\pi_1(S^1)$ topology.
And within the same topology both D-type and N-type vortices
exist. In particular, there are infinitely many D-type
vortices classified by the natural number $k$ which have the same topology.
Moreover, the magnetic flux of these non-Abrikosov vortices
can be integral or fractional, and the integral flux vortex has
the $\pi_2(S^2)$ topology and the fractional flux vortex has
the $\pi_1(S^1)$ topology.}
We show that the N-type vortex
has a $2\pi n/g$-flux or a fractional flux (a fraction of
$2\pi n/g$), but the D-type vortex has $2\pi k/g$ more
flux than the N-type vortex.
These characteristic features of the non-Abrikosov vortex
are clearly absent in the Abrikosov vortex
which carries $2\pi n/g$-flux whose topology is fixed by
$\pi_1(S^1)$.

Next, we show that the non-Abrikosov vortex
can be twisted to form a helical vortex which is periodic
in $z$-axis. {\it More importantly, we show that
we can construct a stable magnetic knot in two-gap superconductors
by smoothly bending the helical vortex and
connecting the periodic ends together. The vortex ring
acquires the knot topology $\pi_3(S^2)$ which is fixed by
the Chern-Simon index of the electromagnetic potential.}
Because of the helical structure of the
magnetic flux the knot has two magnetic flux linked together,
one around the knot tube and one along the knot,
whose linking number is given by the knot quantum number.
And the flux trapped inside the vortex ring provides
a stabilizing repulsive force which prevents the collapse
of the knot, because it can not be squeezed out.
This means that the knot has dynamical (as well as topological)
stability.

It is well-known that multi-gap superconductor may have
interband Josephson interaction \cite{maz}.
We consider a most general quartic Josephson interaction
in two-gap superconductor, and show that the presence of
the Josephson interaction does not affect the existence of the
above topological objects, but can alter the shape of the
solutions drastically. {\it We show that in the presence of
the Josephson interaction we have a magnetic vortex
which can be viewed as a bound state of two fluxes,
which becomes a braided magnetic vortex when twisted.}

The paper is organized as follows. In Section II we discuss
a most general quartic potential in Ginzburg-Landau theory of
two-gap superconductor in mean field approximation, and
study the vacuum structure. In Section III we show that
the Ginzburg-Landau theory of two-gap superconductor
can be understood as a theory of $CP^1$ field coupled to
a scalar field and the electromagnetic field, and argue that
the topology of the theory can be described by the $CP^1$ field
and the electromagnetic field. In Section IV we
construct the non-Abrikosov vortex in two-gap superconductor,
and show that there are two types of boundary condition
which allow two types of magnetic vortex,
the D-type which has no concentration of condensate at the core
and the N-type which has a non-trivial concentration of condensate
at the core. Moreover we show that there are two types of topology,
non-Abelian $\pi_2(S^2)$ and Abelian $\pi_1(S^1)$,
which describes these vortices.
We show that the magnetic flux of these vortices can be
integral or fractional depending on
the parameters of the potential, but the D-type
vortex has $2\pi k/g$ more flux than the N-type vortex.
In Section V we show that we can construct a
helical magnetic vortex in two-gap superconductor,
by twisting the non-Abrikosov vortex and making it
periodic in $z$-axis.
In Section VI we construct the magnetic knot bending
the helical vortex and connecting the periodic ends
together, and show that the knot topology $\pi_3(S^2)$
is described by the Chern-Simon index of the
electromagnetic potential.
In Section VII we consider the Josephson interaction,
and show that the inclusion of the Josephson interaction
does not affect the existence of the topological objects in
two-gap superconductor but alter the shape of the solutions
drastically. In Section VIII we
discuss the non-Abelian superconductivity which can describe
a two-gap superconductor made of two condensates which carry
opposite charge, and argue that the non-Abelian superconductivity
can be realized in liquid metallic hydrogen (LMH).
Finally in Section IX we discuss the physical implications of
our results, and discuss how one can identify these topological
objects in $\rm MgB_2$ and LMH.

\section{Effective potential of Two-gap Superconductor}

In mean field approximation
the free energy of the two-gap
superconductor could be expressed by \cite{baba1,maz,baba2}
\bea
&{\cal H} = \dfrac{\hbar^2}{2m_1} |(\mbox{\boldmath $\nabla$}
+ ig \mbox{\boldmath $A$}) \tilde \phi_1|^2
+\dfrac{\hbar^2}{2m_2} |(\mbox{\boldmath $\nabla$}
+ ig \mbox{\boldmath $A$}) \tilde \phi_2|^2 \nn\\
&+ \tilde V(\tilde \phi_1,\tilde \phi_2)
+ \dfrac{1}{2} (\mbox{\boldmath $\nabla$}
\times \mbox{\boldmath $A$})^2,
\label{scfe1}
\eea
where $\tilde V$ is the effective potential.
We choose the potential to be the most general quartic
potential which has the $U(1)\times U(1)$ symmetry,
\bea
&\tilde V =\dfrac{\tilde{\lambda}_{11}}2|\tilde{\phi}_1|^4
+\tilde{\lambda}_{12}|\tilde{\phi}_1|^2|\tilde{\phi}_2|^2
+\dfrac{\tilde{\lambda}_{22}}2| \tilde{\phi}_2|^4 \nn\\
&-\tilde{\mu}_1|\tilde{\phi}_1|^2
-\tilde{\mu}_2|\tilde{\phi}_2|^2,
\label{scpot1}
\eea
where $\tilde{\lambda}_{ij}$ are the quartic coupling
constants and $\tilde \mu_i$ are the chemical potentials of
$\tilde \phi_i~(i=1,2)$. One might like to include the
Josephson interaction to the potential which breaks the
$U(1)\times U(1)$ symmetry down to $U(1)$.
The Josephson interaction will be discussed separately
in the following. But as we will see, the inclusion of the
Josephson interaction does not alter the qualitative features of
the topological objects we discuss in this paper.

With the normalization
of $\tilde \phi_1$ and $\tilde \phi_2$ to $\phi_1$ and $\phi_2$,
\bea
\phi_1=\dfrac \hbar {\sqrt{2m_1}}\tilde{\phi}_1,\;\;\;\;
\phi_2=\frac \hbar {\sqrt{2m_2}}\tilde{\phi}_2.
\eea
one can simplify the above Hamiltonian (\ref{scfe1}) to
\bea
&{\cal H} = |(\mbox{\boldmath $\nabla$}
+ ig \mbox{\boldmath $A$}) \phi|^2 + V(\phi_1,\phi_2) \nn\\
&+ \dfrac{1}{2} (\mbox{\boldmath $\nabla$}
\times \mbox{\boldmath $A$})^2,
\label{scfe2}
\eea
where $V$ is the normalized potential,
\bea
&V =\dfrac{\lambda_{11}}{2}|\phi_1|^4+\lambda_{12}|\phi_1|^2
|\phi_2|^2+\dfrac{\lambda_{22}}{2}|\phi_2|^4 \nn\\
&-\mu_1|\phi_1|^2-\mu_2|\phi_2|^2 \nn\\
&= \dfrac{\lambda_{11}}{2} \big( |\phi_1|^2
-\hat \phi_1^2 \big)^2 +\dfrac{\lambda_{22}}{2}\big( |\phi_2|^2
- \hat \phi_2^2 \big)^2 \nn\\
&+\lambda_{12} \big(|\phi_1|^2 -\hat \phi_1^2 \big)
\big(|\phi_2|^2 -\hat \phi_2^2 \big) + V_0, \nn\\
&\hat \phi_1^2 = \dfrac{\mu_1\lambda_{22}-\mu_2\lambda_{12}}\Delta,
~~~~~\hat \phi_2^2 = \dfrac{\mu_2\lambda_{11}
-\mu_1\lambda_{12}}\Delta \nn\\
&V_0 = -\dfrac{\lambda_{11} \mu_2^2
+\lambda_{22} \mu_1^2 -2 \lambda_{12} \mu_1 \mu_2}{2\Delta}, \nn\\
&\Delta = \lambda_{11} \lambda_{22} - \lambda_{12}^2.
\label{scpot2}
\eea
Notice that the potential can also be written as
\bea
&V = \dfrac{1}{2\lambda_{11}} \Big[(\lambda_{11} |\phi_1|^2
+\lambda_{12} |\phi_2|^2 -\mu_1)^2 \nn\\
&+ \Delta \big( |\phi_2|^2 -\hat \phi_2^2 \big)^2 \Big] +  V_0 \nn\\
&= \dfrac{1}{2\lambda_{22}} \Big[(\lambda_{12} |\phi_1|^2
+\lambda_{22} |\phi_2|^2 -\mu_2)^2 \nn\\
&+ \Delta \big( |\phi_1|^2 -\hat \phi_1^2 \big)^2 \Big] +  V_0.
\label{scpot3}
\eea
From now on we will assume that all coupling constants
except $\lambda_{12}$ are positive.

To find the vacuum of the potential let
\bea
&\dfrac{\partial V}{\partial |\phi _1|^2}=\lambda _{11}|\phi _1|^2
+\lambda_{12}|\phi _2|^2-\mu _1=0, \nn\\
&\dfrac{\partial V}{\partial |\phi _2|^2}=\lambda _{12}|\phi _1|^2
+\lambda_{22}|\phi _2|^2-\mu _2=0,
\label{vcon}
\eea
and find the extremum
\bea
&|\phi _1|^2=\hat \phi_1^2,
~~~~~|\phi_2|^2= \hat \phi_2^2.
\label{vac}
\eea
To check whether this extremum is the maximum or minimum,
consider the Hessian
\bea
&\det H=\det \dfrac{\partial^2V}
{\partial |\phi _i|^2\partial |\phi _j|^2}
=\lambda _{11}\lambda _{22}-\lambda _{12}^2 \nn\\
&=\Delta.
\label{hess}
\eea
There are three possibilities; positive,
zero, or negative $\Delta$. We consider each case separately.

A. $\Delta >0$: In this case we have
$\lambda_{12}^2<\lambda_{11}\lambda_{22}$,
and the extremum (\ref{vac}) becomes the local minimum.
But since $|\phi _i|$ have to be positive we have
the following vacuum
\bea
&\Bigg( \matrix{<|\phi_1|> \cr <|\phi_2|> } \Bigg)
=\Bigg( \matrix{\hat \phi_1
\cr \hat \phi_2 }\Bigg),
\label{vaca1}
\eea
for $\lambda_{12} \leq 0$ or for
\bea
&0< \lambda_{12},~~~\dfrac{\lambda_{12}}{\lambda_{22}}
<\dfrac{\mu _1}{\mu _2}
<\dfrac{\lambda _{11}}{\lambda _{12}}.
\label{vcona1}
\eea
Notice that both $<|\phi_1|>$ and $<|\phi_2|>$ are
non-vanishing. But for
\bea
&0< \lambda_{12},~~~\dfrac{\lambda_{12}}{\lambda_{22}}
<\dfrac{\lambda_{11}}{\lambda_{12}} <\dfrac{\mu_1}{\mu_2},
\label{vcona21}
\eea
we have the following vacuum from (\ref{scpot3}),
\bea
&\Bigg( \matrix{<|\phi_1|> \cr <|\phi _2|> } \Bigg)
=\Bigg( \matrix{\sqrt{\mu_1/\lambda_{11}} \cr 0 }\Bigg),
\label{vaca2}
\eea
Finally, when
\bea
&0< \lambda_{12},~~~\dfrac{\mu_1}{\mu_2}
< \dfrac{\lambda_{12}}{\lambda_{22}}
<\dfrac{\lambda_{11}}{\lambda_{12}},
\label{vcona22}
\eea
we can always transform this case to the case
(\ref{vcona21}) by re-labeling $\phi_1$ and
$\phi_2$ as $\phi_2$ and $\phi_1$, so that in this case
we can assume that the vacuum is still given by
(\ref{vaca2}) without loss of generality.

B. $\Delta =0$: In this case we have
$\lambda_{12}^2=\lambda_{11} \lambda_{22}$,
and the potential (\ref{scpot2}) is reduced to
\bea
&V =\dfrac{1}{2\lambda_{11}} \Big[ \big( \lambda_{11}|\phi_1|^2
+\lambda_{12} |\phi_2|^2 -\mu_1 \big)^2 \nn\\
&-2(\mu_2 \lambda_{11} -\mu_1 \lambda_{12}) |\phi_2|^2 \Big] \nn\\
&=\dfrac{1}{2\lambda_{22}} \Big[ \big(\lambda_{12}|\phi_1|^2
+\lambda_{22} |\phi_2|^2 -\mu_2 \big)^2 \nn\\
&-2(\mu_1 \lambda_{22} -\mu_2 \lambda_{12}) |\phi_1|^2 \Big].
\label{scpot4}
\eea
So for $\lambda_{12} <0$, the potential becomes unbounded from below,
so that it has no minimum. For
\bea
0< \lambda_{12},
~~~~~\dfrac{\lambda_{11}}{\lambda_{12}}
=\dfrac{\lambda_{12}}{\lambda_{22}}<\dfrac{\mu_1}{\mu_2},
\label{vconb1}
\eea
we have the following vacuum
\bea
\Bigg( \matrix{ <|\phi_1|> \cr <|\phi_2|> } \Bigg)
=\Bigg( \matrix{ \sqrt{\mu_1/\lambda_{11}} \cr
0  } \Bigg).
\label{vacb1}
\eea
Next, consider the case
\bea
0< \lambda_{12},
~~~~~\dfrac{\mu_1}{\mu_2}< \dfrac{\lambda_{11}}{\lambda_{12}}
=\dfrac{\lambda_{12}}{\lambda_{22}}.
\label{vconb2}
\eea
But this can be transformed to (\ref{vconb1}) with
the re-labeling. So we can assume that the vacuum is given by
(\ref{vacb1}) without loss of generality.
Finally, when
\bea
\dfrac{\mu_1}{\mu_2} =\dfrac{\lambda_{11}}{\lambda_{12}}
=\dfrac{\lambda_{12}}{\lambda_{22}},
\label{vconb3}
\eea
we have the degenerate vacuum
\bea
\mu_1 <|\phi_1|>^2 + \mu_2 <|\phi_2|>^2
= \dfrac{\mu_1 \mu_2}{\lambda_{12}}.
\label{vacb3}
\eea
This case includes the special (and familiar) $SU(2)$ symmetric
case
\bea
&\lambda_{11} =\lambda_{12} = \lambda_{22} = \lambda,
~~~u_1 =\mu_2 =\mu, \nn\\
&<|\phi_1|>^2 + <|\phi_2|>^2 = \dfrac{\mu}{\lambda}.
\label{vacsu2}
\eea
In this case the Hamiltonian (\ref{scfe2}) has the full
$SU(2)$ symmetry.

C. $\Delta <0$: In this case we have
$\lambda_{12}^2>\lambda_{11}\lambda_{22}$,
and the extremum (\ref{vac}) becomes the local maximum.
So the minimum state must satisfy
\bea
|\phi_1|^2|\phi_2|^2=0.
\eea
Now, by inspection one can show that when
\bea
0< \lambda_{12}, ~~~\dfrac{\lambda_{11}}{\lambda_{12}}
<\sqrt{\dfrac{\lambda_{11}}{\lambda_{22}}}
<\dfrac{\mu_1}{\mu_2},
\label{vconc1}
\eea
the vacuum must be
\bea
&\Bigg( \matrix{<|\phi_1|> \cr <|\phi_2|>} \Bigg)
= \Bigg( \matrix{\sqrt{\mu_1/\lambda_{11}} \cr 0 } \Bigg).
\label{vacc1}
\eea
Next, consider the case
\bea
\dfrac{\mu_1}{\mu_2}<\sqrt{\dfrac{\lambda_{11}}{\lambda_{22}}}
<\dfrac{\lambda_{12}}{\lambda_{22}}.
\eea
This case can be reduced
to the above case (by re-labeling $\phi_1$ and $\phi_2$),
so that when $\lambda_{12}$ is positive
one can assume that the vacuum is given by (\ref{vacc1})
without loss of generality.
Finally when $\lambda_{12}$ is negative the potential
has no minimum, because it is unbounded from below.
This must be clear from (\ref{scpot4}).

In summary, we have three types of vacuum state: \\
A. Type I: Integer flux vacuum
\bea
\Bigg( \matrix{ <|\phi_1|> \cr <|\phi_2|> } \Bigg)
=\Bigg( \matrix{ \sqrt{\mu_1/\lambda_{11}} \cr
0  } \Bigg).
\label{vacb}
\eea
This is possible when we have one of the following
three cases,
\bea
&&(a)~~~~0< \lambda_{12},~~~\dfrac{\lambda_{12}}{\lambda_{22}}
<\dfrac{\lambda_{11}}{\lambda_{12}} \leq \dfrac{\mu_1}{\mu_2}, \nn\\
&&(b)~~~~0< \lambda_{12},~~~\dfrac{\lambda_{11}}{\lambda_{12}}
<\sqrt{\dfrac{\lambda_{11}}{\lambda_{22}}}
<\dfrac{\mu_1}{\mu_2}, \nn\\
&&(c)~~~~0< \lambda_{12},~~~\dfrac{\lambda_{11}}{\lambda_{12}}
=\dfrac{\lambda_{12}}{\lambda_{22}}<\dfrac{\mu_1}{\mu_2}.
\eea
We call this integer flux vacuum because, as we will see,
for this type of vacuum the magnetic vortex has
an integer flux. \\
B. Type II: Fractional flux vacuum
\bea
\left( \begin{array}{l}
<|\phi_1|> \\ <|\phi_2|> \end{array} \right)
=\left( \begin{array}{l} \hat{\phi}_1 \\ \hat{\phi}_2
\end{array} \right).
\label{vaca}
\eea
This is possible when we have one of the following
three cases,
\bea
&&(a)~~~~\lambda_{12} <0,~~~\dfrac{|\lambda_{12}|}{\lambda_{22}}
<\dfrac{\lambda_{11}}{|\lambda_{12}|}, \nn\\
&&(b)~~~~0< \lambda_{12},~~~\dfrac{\lambda_{12}}{\lambda_{22}}
<\dfrac{\mu_1}{\mu_2}<\dfrac{\lambda_{11}}{\lambda_{12}}, \nn\\
&&(c)~~~~\lambda_{12} =0.
\eea
We call this fractional flux vacuum because, as we will see,
for this type of vacuum the magnetic vortex has
a fractional flux. \\
C. Type III: Degenerate vacuum
\bea
\mu_1 <|\phi_1|>^2 + \mu_2 <|\phi_2|>^2 = \dfrac{\mu_1 \mu_2}{\lambda_{12}}.
\label{vacc}
\eea
This is what we have when
\bea
\dfrac{\mu_1}{\mu_2}=\dfrac{\lambda_{12}}{\lambda_{11}}
=\dfrac{\lambda_{22}}{\lambda_{12}}.
\eea
Notice that the potential (\ref{scpot2}) has no vacuum
when
\bea
\lambda_{12}< 0,~~~\dfrac{\lambda_{11}}{|\lambda_{12}|}
\leq \dfrac{|\lambda_{12}|}{\lambda_{22}}.
\eea
All other cases can be reduced to one of the above cases
by re-labelling $\phi_1$ and $\phi_2$.
As we will see the vacuum structure will play an important
role in the following.

Notice that with
\bea
&\lambda =\dfrac{\lambda_{11}+\lambda _{22}+2\lambda _{12}}4, \nn\\
&\alpha =\dfrac{\lambda_{11}-\lambda_{22}}2,
~~~\beta =\dfrac{\lambda_{11}+\lambda_{22}-2\lambda_{12}}4, \nn\\
&\mu =\dfrac{\mu_1+\mu_2}2,~~~\gamma =\dfrac{\mu _1-\mu _2}2,
\eea
we have
\bea
&\lambda_{11}=\lambda+\beta+\alpha,
~~~~~\lambda_{22}=\lambda+\beta-\alpha,  \nn\\
&\lambda_{12}=\lambda-\beta, \nn\\
&\mu_1=\mu+\gamma,~~~~\mu_2=\mu-\gamma,
\eea
so that the potential (\ref{scpot2}) can be written as
\bea
&V= \dfrac {\lambda} 2(|\phi_1|^2+|\phi_2|^2-\dfrac{\mu}\lambda)^2
+\dfrac {\alpha}2 (|\phi_1|^4-|\phi_2|^4) \nn\\
&+\dfrac {\beta}2 (|\phi_1|^2-|\phi_2|^2)^2
-\gamma (|\phi_1|^2-|\phi_2|^2)-\dfrac{\mu^2}{2\lambda}.
\label{scpot5}
\eea
In terms of the new parameters we have
\bea
&\hat \phi_1^2= \dfrac {2 (\beta\mu+\gamma\lambda)
-\alpha(\mu+\gamma)}{\Delta}, \nn\\
&\hat \phi_2^2= \dfrac {2 (\beta\mu-\gamma\lambda)
+\alpha(\mu-\gamma)}{\Delta},
\eea
where $\Delta=4 \beta\lambda-\alpha^2$.

\section{Dynamics of two-gap condensates}

With this preliminary one may study the topological objects
of two-gap superconductor minimizing the free energy.
On the other hand, to study a static solution,
one might as well start from the following relativistic
Ginzburg-Landau Lagrangian
\bea
&{\cal L} = - |D_\mu \phi|^2 + V(\phi)
- \dfrac{1}{4} F_{\mu \nu}^2, \nn\\
&D_\mu \phi = (\partial_\mu + ig A_\mu) \phi,
\label{sclag}
\eea
which reproduces the the free energy (\ref{scfe2}) in
the static limit.
The Lagrangian has the equation of motion
\bea
&D^2\phi_1 = \dfrac{\partial V}{\partial |\phi_1|^2} \phi_1, \nn\\
&D^2\phi_2 = \dfrac{\partial V}{\partial |\phi_2|^2} \phi_2, \nn\\
&\partial_\mu F_{\mu \nu} = j_\nu =  i g \Big[(D_\nu
\phi)^{\dagger}\phi - \phi ^{\dagger}(D_\nu \phi) \Big].
\label{sceq1}
\eea
To understand the meaning of this we let
\bea
&\phi =\dfrac{1}{\sqrt 2} \rho \xi,
~~~~~|\phi^{\dagger}\phi|=\dfrac{\rho}{2},
~~~~~{\xi}^{\dagger}\xi = 1, \nn\\
&\hat n = \xi^{\dagger} \vec \sigma \xi,
\label{ndef}
\eea
and find the following identities
\bea
& (\pro_\mu \hn)^2 = 4 \big(|\pro_\mu \xi|^2
-|\xi^\dag \pro_\mu \xi|^2 \big), \nn\\
& -\dfrac{1}{g} \hn \cdot (\pro_\mu \hn \times \pro_\nu \hn)
= \dfrac{2i}{g} (\pro_\mu \xi^\dag \pro_\nu \xi
- \pro_\nu \xi^\dag \pro_\mu \xi ) \nn\\
&= \pro_\mu C_\nu - \pro_\nu C_\mu, \nn\\
&\Big[\pro_\mu  +\dfrac{1}{2} (ig C_\mu - \vec \sigma
\cdot \pro_\mu \n) \Big] \xi =0, \nn\\
&C_\mu =\dfrac{2i}g\xi ^{\dagger }\partial _\mu \xi.
\label{id}
\eea
From these we can reduce (\ref{sceq1}) to \cite{ijpap,prb05}
\begin{eqnarray}
&\partial ^2\rho -\Big( \dfrac 14(\partial _\mu \hn)^2+g^2(A_\mu
-\dfrac 12C_\mu )^2\Big) \rho   \nonumber \\
&=\Big[\dfrac \lambda 2 (\rho ^2- \bar \rho^2)
+\big(\dfrac \alpha 2\rho^2-\gamma \big) n_3
+\dfrac \beta 2\rho^2 n_3^2 \Big] \rho, \nonumber \\
&\hat{n}\times \partial ^2\hat{n}+2\dfrac{\partial _\mu \rho }%
\rho \hat{n}\times \partial _\mu \hat{n}-\dfrac 2{g\rho^2}
\partial_\mu F_{\mu \nu }\partial_\nu \hat{n}  \nonumber \\
&=\Big(2\gamma -(\dfrac{\alpha}2 + \beta n_3) \rho^2 \Big)
\hat{k}\times \hat{n},  \nonumber \\
&\partial_\mu F_{\mu \nu }=j_\nu =g^2\rho ^2\Big(A_\nu
-\dfrac12 C_\nu \Big),  \nonumber \\
&\bar \rho^2=\dfrac{2\mu}\lambda.
\label{sceq2}
\end{eqnarray}
This is the equation for two-gap superconductor,
which allows a large class of interesting topological
objects, straight magnetic vortex, helical magnetic vortex,
and magnetic knot, all with $4\pi/g$-flux, $2\pi/g$-flux,
or fractional flux.

The equation (\ref{sceq1}) is an equation of the complex doublet
$\phi$ which has four degrees. But notice that the equation
(\ref{sceq2}) is, except for $C_\mu$, expressed
completely in terms of the $CP^1$ field $\hn$ and
the scalar field $\rho$. Moreover,
(\ref{id}) tells that $C_\mu$ can also be written in terms of
$\hn$. In fact $\hn$ uniquely defines a righthanded
orthonormal frame ($\hat n_1,\hat n_2,\hat n$), with
$\hat n_1 \times \hat n_2=\hat n$, up to the $U(1)$ rotation
which leaves $\hat n$ invariant. Then $C_\mu$ is given (up to a $U(1)$
gauge transformation) by the Mermin-Ho relation \cite{ho,cho79,cho80,cho81}
\bea
&C_\mu = -\dfrac{1}{g} \hat n_1 \cdot \pro_\mu \hat n_2, \nn\\
&\pro_\mu C_\nu - \pro_\nu C_\mu
=-\dfrac{1}{g} \hn \cdot (\pro_\mu \hn \times \pro_\nu \hn).
\eea
This tells that we can transform the equation (\ref{sceq1})
of the complex doublet condensate $\phi$ to
the equation (\ref{sceq2}) of the $CP^1$ field $\hn$
and the scalar field $\rho$. In fact, with
\bea
&B_\mu= A_\mu-\dfrac12 C_\mu, \nn\\
&G_{\mu\nu}= \pro_\mu B_\nu - \pro_\nu B_\mu,
\label{bm}
\eea
we can express (\ref{sceq2}) completely in terms of $\hn$,
$\rho$, and $B_\mu$. This is not accidental.
Indeed with (\ref{ndef}), (\ref{id}), and (\ref{bm}),
we can express the Hamiltonian (\ref{scfe2}) as
\bea
&{\cal H}=\dfrac 12 (\partial_\mu \rho)^2 + \dfrac 12 g^2\rho^2 B^2_\mu
+\dfrac 18\rho^2(\partial_\mu \hat n)^2 + V(\rho,n_3) \nn\\
&+\dfrac 14 \big[G_{\mu\nu}-\dfrac {1}{2g} \hat n
\cdot (\partial_\mu \hat n \times \partial_\nu \hat n)  \big]^2,
\label{scfe3}
\eea
where
\bea
&V(\rho,n_3)=\dfrac \lambda 8 \big(1 +\dfrac{\alpha}\lambda n_3
+\dfrac{\beta}\lambda  n_3^2 \big) \rho^4 \nn\\
&-\dfrac\mu 2 \big(1+\dfrac{\gamma}\mu n_3 \big) \rho^2,
\eea
and $n_3 = \xi_1^* \xi_1 - \xi_2^* \xi_2$.
This means that the Ginzburg-Landau
theory of two-gap superconductor can be understood
as a theory of $CP^1$ field $\hn$ (coupled to $\rho$
and $B_\mu$) \cite{ijpap,prb05,baba1,cm2}.
This is because the $U(1)$ gauge invariance of (\ref{sclag})
reduces the physical degrees
of the complex doublet $\phi$ to $\rho$ and $\hn$, and
the massive photon $B_\mu$.
As we will see, this has a very important physical implication,
because this tells that the topology of two-gap superconductor
can be described by the topology of $\hn$ and $A_\mu$.

The equation (\ref{sceq2}) allows two conserved currents,
the electromagnetic current $j_\mu$ and the
neutral current $k_\mu$ \cite{cm2},
\bea
&j_\mu = g^2 \rho^2(A_\mu -\dfrac 12C_\mu) \nn\\
&k_\mu = g^2 \rho^2 \Big[ A_\mu \big(\xi_1^{*}\xi_1
-\xi_2^{*}\xi_2 \big)
+\dfrac{i}{g} \big(\partial_\mu \xi_1^{*}\xi_1 \nn\\
&+\xi_2^{*}\partial_\mu \xi_2 \big) \Big],
\label{2sc1}
\eea
which are nothing but the Noether
currents of the $U(1)\times U(1)$ symmetry of the Hamiltonian
(\ref{scfe2}). Indeed they are
the sum and difference of two electromagnetic currents of
$\phi_1$ and $\phi_2$
\bea
j_\mu = j^{(1)}_\mu + j^{(2)}_\mu,
~~~~~k_\mu = j^{(1)}_\mu - j^{(2)}_\mu.
\label{2sc2}
\eea
Clearly the conservation of $j_\mu$ follows from
the last equation of (\ref{sceq2}). But the conservation of
$k_\mu$ comes from the second equation of (\ref{sceq2}),
which (together with the last equation) tells the
existence of a partially conserved $SU(2)$ current
$\vec j_\mu$ \cite{cm2},
\bea
\vec j_\mu = g\rho^2 \Big(\frac 12\vec{n} \times \partial_\mu
\vec{n} -g(A_\mu -\frac 12C_\mu )\vec{n} \Big).
\eea
This $SU(2)$ current is exactly conserved when
$\alpha=\beta=\gamma=0$. But notice that
$k_\mu=\hat k \cdot \vec j_\mu$.
This assures that we have the conservation of $k_\mu$
even when $\alpha \beta \gamma \neq 0$. It is interesting to notice
that $j_\mu$ and $k_\mu$ are precisely the $\hn$ and $\hat k$
components of $\vec j_\mu$.

\section{Non-Abrikosov Vortex}

The two-gap superconductor allows different types of
interesting magnetic vortex \cite{ijpap,prb05,cm2,baba2}.
In terms of the structure of the vortex
they are classified to two types, the D-type vortex
which has no concentration of the condensates at the core
and the N-type which has a non-vanishing concentration of
the condensates at the core.
The reason for this is that, unlike
the Abrikosov vortex in ordinary superconductor,
the vortex in two-gap superconductor allows two different
boundary conditions at the core.

To discuss the straight vortex let
$(\varrho,\varphi,z)$ be the cylindrical coordinates and
choose the ansatz
\begin{eqnarray}
&\rho =\rho (\varrho ),~~~~~\xi =\left(
\begin{array}{c}
\cos \dfrac{f(\varrho )}2\exp (-in\varphi ) \\
\sin \dfrac{f(\varrho )}2
\end{array} \right) , \nn\\
&A_\mu =\dfrac ngA(\varrho)\partial_\mu \varphi.
\label{svans}
\end{eqnarray}
With this we have
\bea
&\hat{n}=\xi ^{\dag }\vec{\sigma}\xi =\left(
\begin{array}{c}
\sin f(\varrho )\cos n\varphi  \\
\sin f(\varrho )\sin n\varphi  \\
\cos f(\varrho )
\end{array} \right),  \nonumber \\
&C_\mu =n \dfrac{\cos {f(\varrho )}+1}g \partial_\mu \varphi, \nn\\
& j_\mu =ng\rho^2\Big(A-\dfrac{\cos {f}+1}{2} \Big)
 \partial_\mu \varphi,  \nn\\
& k_\mu = ng\rho^2\Big(A \cos f-\dfrac{\cos {f}+1}{2} \Big)
\partial_\mu \varphi,
\label{sc}
\eea
and the Hamiltonian (\ref{scfe3}) becomes
\begin{eqnarray}
&{\cal H} =\dfrac 12\dot{\rho}^2+\dfrac 18\rho
^2 \Big(\dot{f}^2+\dfrac{n^2}{\varrho^2}\sin ^2f \Big)  \nn\\
&+ \dfrac{n^2\rho
^2}{2\varrho^2}\Big( A - \dfrac {\cos f+1}{2} \Big) ^2
+\dfrac{n^2}{2g^2\varrho^2}\dot{A}^2 \nn \\
&+\dfrac \lambda 8  \Big[(\rho ^2-\bar \rho^2)^2
+\dfrac{\alpha}\lambda (\rho^2-\dfrac{4\gamma }\alpha) \rho^2 \cos f \nn\\
&+\dfrac{\beta}\lambda \rho^4\cos ^2f \Big]
-\dfrac{\mu^2}{2\lambda}.
\label{scfe4}
\end{eqnarray}
With this (\ref{sceq2}) becomes
\begin{eqnarray}
&\ddot{\rho}+\dfrac{1}{\varrho}\dot{\rho}-\Big[\dfrac 14
\Big(\dot{f}^2+\dfrac{n^2}{\varrho^2}\sin^2f\Big) \nn\\
&+\dfrac{n^2}{\varrho^2}
\Big(A-\dfrac{\cos {f}+1}{2} \Big)^2\Big]\rho   \nonumber \\
&=\dfrac \lambda 2  \Big[(\rho^2- \bar \rho^2)
+\dfrac{\alpha}\lambda (\rho^2-\dfrac{2\gamma }\alpha)\cos f
+\dfrac{\beta}\lambda \rho^2\cos ^2f \Big]\rho,  \nonumber \\
&\ddot{f}+\Big(\dfrac{1}{\varrho}+2\dfrac{\dot{\rho}}\rho %
\Big)\dot{f}-2\dfrac{n^2}{\varrho^2}\Big(A-\dfrac 12\Big)%
\sin f  \nonumber \\
&=\Big(2\gamma -(\dfrac \alpha 2
+\beta \cos f)\rho^2 \Big) \sin f,  \nonumber \\
&\ddot{A}-\dfrac{1}{\varrho}\dot{A}-g^2\rho ^2\Big(A-\dfrac{\cos {f}+1}{2} \Big)=0.
\label{sveq}
\end{eqnarray}
Notice that this can also be derived by minimizing the
Hamiltonian (\ref{scfe4}).

To solve the equation, we have to fix the boundary
conditions. To determine the possible boundary condition
at the core we expand $\rho (\varrho)$, $f(\varrho)$,
and $A(\varrho)$ near the origin as
\begin{eqnarray}
&\rho (\varrho) \simeq \rho _0+\rho _1\varrho
+\rho _2\varrho^2+\rho _3\varrho^3+..., \nn\\
&f(\varrho) \simeq f_0+f_1\varrho+f_2\varrho^2+f_3\varrho^3+..., \nn\\
&A(\varrho) \simeq a_0+a_1\varrho+a_2\varrho^2+a_3\varrho^3+...,
\end{eqnarray}
and find that the smoothness at the core requires
\begin{widetext}
\bea
&\dfrac{\rho_0}4 \Big( (2a_0- \cos f_0 -1)^2+\sin^2 f_0 \Big)
\dfrac{1}{\varrho^2} + \Big[ \rho_0 \Big( \dfrac12 (2a_0-1)
f_1 \sin f_0 + (2 a_0 -\cos f_0-1) a_1 \Big) \nn\\
&+\dfrac{\rho_1}4 \Big((2a_0- \cos f_0 -1)^2+\sin^2 f_0
-\dfrac{4}{n^2} \Big) \Big] \dfrac1\varrho...
+\Big[ ...+ \dfrac{\rho_k}4 \Big((2a_0- \cos f_0 -1)^2+\sin^2 f_0
-\dfrac{4 k^2}{n^2} \Big) \Big] \varrho^{k-2}+...=0, \nn\\
&\left \{ \matrix {\dfrac{(2a_0-1)\sin f_0}{\varrho^2}
+\Big(\big((2a_0-1)\cos f_0-\dfrac{1}{n^2} \big) f_1 +2a_1\sin f_0\Big)
\dfrac 1\varrho +...=0,~~~\rho_0\neq 0  \cr
\dfrac{(2a_0-1)\sin f_0}{\varrho^2}
+\Big(\big((2a_0-1)\cos f_0-\dfrac{3}{n^2} \big) f_1 +2a_1\sin f_0\Big)
\dfrac 1\varrho +...=0,~~~\rho_0=0,~\rho_1\neq 0 \cr
.  \cr .  \cr .  \cr
\dfrac{(2a_0-1)\sin f_0}{\varrho^2}
+\Big(\big((2a_0-1)\cos f_0-\dfrac{2k+1}{n^2} \big) f_1 +2a_1\sin f_0\Big)
\dfrac 1\varrho +...=0,~~~\rho_0=\rho_1=...\rho_{k-1}=0,
~\rho_k\neq 0 } \right. \nn\\
&\dfrac{a_1}\varrho+\dfrac{g^2\rho_0^2}{2}
\big(2a_0-\cos f_0 -1 \big)
-\Big(3a_3-g^2\rho_0 \rho_1(2a_0-\cos f_0-1) \nn\\
&-g^2\rho_0^2(a_1+\dfrac 12 f_1\sin f_0) \Big) \varrho+...=0.
\label{sccbc}
\eea
\end{widetext}
Now, consider the case $\rho_0\neq 0$ first.
In this case the last equation
requires $a_1=0$ and $2a_0-1=\cos f_0$,
and the first equation requires
$\sin f_0=0$ and $\rho_1=0$. With $\sin f_0=0$
we must either have $f_0=\pi$ and $a_0=0$ or $f_0=0$ and $a_0=1$.
And with $f_0=\pi$, we have $f_1=0$
for $n \neq \pm 1$ from the second equation.
One might choose $f_0=0$ instead of
$f_0=\pi$, but this does not lead us to a new solution.
Next, consider the case $\rho_0=0,~\rho_1 \neq 0$.
In this case the first two sets of equations tells
that we may have $f_0=\pi$, $a_0=\pm 1/n$, and $f_1=0$
for $n\neq \mp 1$ or $n\neq \pm 3$.
Similarly, for $\rho_0=\rho_1=...=\rho_{k-1}=0,~\rho_k \neq 0$,
(\ref{sccbc}) tells that we may have $f_0=\pi$, $a_0=\pm k/n$,
and $f_1=0$ for $n\neq \mp 1$ or $n\neq \pm (2k+1)$.
This tells that we can choose the following boundary condition at
the core for the vortex described by the ansatz (\ref{svans})
in two-gap superconductor \cite{ijpap,prb05,cm2}: \\
A. Dirichlet boundary condition
\bea
&\rho(0)=0,~~~\dot{\rho}(0) \neq 0, \nn\\
&A(0)=-\dfrac{1}{n},~~~f(0)=\pi,~~~\dot f=0 ~for ~n \neq 1.
\label{dbc}
\eea
In particular, for $n=1$ we must have $A(0)=-1$.
In general, the Dirichlet boundary
condition can be written as
\bea
&\rho(0)=\dfrac{d \rho}{d \varrho}(0)
=...=\dfrac{d^{k-1} \rho}{d \varrho^{k-1}}(0)=0,
~~~\dfrac{d^k \rho}{d \varrho^k}(0) \neq 0, \nn\\
&A(0)=-\dfrac{k}{n},~~~f(0)=\pi,~~~\dot f=0 ~for ~n \neq 1,
\label{dbcg}
\eea
so that for $n=1$ we must have $A(0)=-k$. \\
B. Neumann boundary condition
\bea
&\rho(0)\neq 0,~~~\dot{\rho}(0)=0, \nn\\
&A(0)=0,~~~f(0)=\pi,~~~\dot f=0 ~for ~n \neq 1.
\label{nbc}
\eea
So here we have $A(0)=0$ for all $n$.
This shows that the magnetic vortex in two-gap superconductor
allows two types of boundary condition which are different
from what we have in ordinary superconductor.
This is a new feature of two-gap superconductor
which will have a deep impact in the following.

To determine the boundary condition at the infinity
notice that at the infinity all fields must assume the
vacuum values. In particular, the electromagnetic current
must vanish at the infinity.
This means that we must have
\begin{eqnarray}
&\rho(\infty)= \sqrt{2(<|\phi_1|>^2+<|\phi_2|>^2)},\nn\\
&\cos f(\infty)=<n_3>=\dfrac{<|\phi_1|>^2-<|\phi_2|>^2}
{<|\phi_1|>^2+<|\phi_2|>^2}, \nn\\
& A(\infty)=\dfrac {\cos f(\infty)+1}{2} \nn\\
&=\dfrac {<|\phi_1|>^2} {<|\phi_1|>^2+<|\phi_2|>^2}.
\label{bcinf}
\end{eqnarray}
Notice that for the integer flux vacuum we have $A(\infty)=1$,
but for the fractional flux vacuum $A(\infty)$ becomes
fractional.

At this point one might worry about the apparent singularity
at the core in the gauge potential when $A(0) \neq 0$.
But this singularity is a coordinate singularity
which can easily be removed by a gauge transformation.
Indeed one can always choose a gauge where $A(0)$ becomes zero
to remove the coordinate singularity. But notice that
the gauge transformation also changes $A(\infty)$
by the same amount, leaving $A(\infty)-A(0)$ invariant.

\begin{figure}[t]
\includegraphics[scale=0.4]{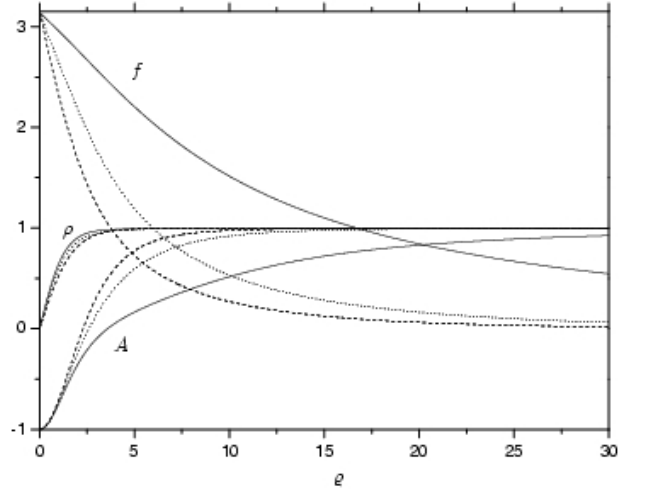}
\caption{The D-type straight vortex with $n=1$ and $k=1$
which has $4\pi/g$ flux.
Three solutions are shown: $\alpha=\beta=\gamma=0$ (solid
lines), $\alpha=\beta=0$, $\gamma =0.005$ (dashed lines),
and $\alpha=\gamma=0$, $\beta =-0.005$ (dotted lines).
Here the unit of the scale is $1/{\bar \rho}$ and
we have put $\lambda/g^2=2$.}
\label{sv4pi}
\end{figure}

The existence of two types of boundary conditions
in two-gap superconductor has an important impact.
To understand this notice that the magnetic flux of vortex
is given by
\begin{eqnarray}
\Phi = \oint A_\mu dx^\mu = \big( A(\infty)-A(0) \big)
\dfrac {2 \pi n}{g}.
\label{mflux}
\end{eqnarray}
Now, it is clear that the magnetic flux becomes fractional
when $A(\infty)$ is fractional, which happens when
$<n_3> \neq 1$ (or equivalently $<|\phi_2|>\neq 0$).
As importantly, when $A(\infty)=1$ the magnetic flux becomes
$2\pi(n+k)/g$ with $A(0)=-k/n$. This was impossible
in ordinary superconductor.
Now we classify the magnetic vortex in terms of the flux.

\subsection{$4\pi/g$-flux vortex}

Let us choose the Dirichlet boundary condition
at the core and the integer flux vacuum
at the infinity \cite{ijpap,cm2}. With $k=1$ we require
\begin{eqnarray}
&\rho (0)=0,~~\dot \rho (0) \neq 0,~~~\rho (\infty )
=\sqrt{\dfrac{2(\mu +\gamma )}{(\lambda +\alpha+\beta )}}, \nn\\
&f(0)=\pi,~~~~~f(\infty ) =0, \nn\\
&A(0)=-\dfrac{1}{n},~~~~~A(\infty )=1.
\label{dbc4pi}
\end{eqnarray}
With this we can integrate (\ref{sveq}) to find the vortex solutions.
The solutions with $n=1$ with different parameters
are shown in Fig.~\ref{sv4pi}.
We call this a D-type vortex, because this comes from
the Dirichlet boundary condition at the core.
Both $\phi_1$ and $\phi_2$ start from zero at the core.
However, notice that $\phi_1$ approaches the finite vacuum value
but $\phi_2$ approaches zero at the infinity. So $\phi_2$
has a maximum concentration at a finite distance
from the core. This is a generic feature of a D-type vortex.

\begin{figure}[t]
\includegraphics[scale=0.4]{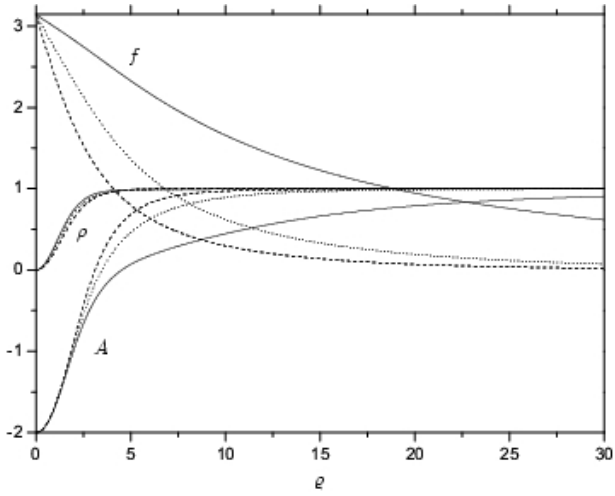}
\caption{The D-type straight vortex with $n=1$ and $k=2$
which has $6\pi/g$ flux.
Three solutions are shown: $\alpha=\beta=\gamma=0$ (solid lines),
$\alpha=\beta=0$, $\gamma =0.005$ (dashed lines), and
$\alpha=\gamma=0$, $\beta =-0.005$ (dotted lines). Here the unit
of the scale is $1/{\bar \rho}$ and we have put $\lambda/g^2=2$.}
\label{sv6pi}
\end{figure}

With (\ref{dbc4pi}) the magnetic flux is given by
\begin{eqnarray}
&\Phi =\dfrac{}{} \int F_{\varrho \varphi}d^2x
=\dfrac{}{} \int \partial_\varrho A_\varphi d^2x \nn\\
&=\Big(1+ \dfrac{1}{n} \Big) \dfrac{2\pi n}g=\dfrac{2\pi}{g}(n+1),
\end{eqnarray}
so that when $n=1$ the vortex has $4\pi/g$ flux.
Moreover, the solution has a non-Abelian topology. To see this notice that
$\hn$ defines a mapping $\pi_2(S^2)=n$ from the compactifed
$xy$-plane $S^2$ to the $CP^1$ space $S^2$.
Clearly this is non-Abelian.

In general we may require
\bea
&\rho(0)=\dfrac{d \rho}{d \varrho}(0)
=...=\dfrac{d^{k-1} \rho}{d \varrho^{k-1}}(0)=0, \nn\\
&\dfrac{d^k \rho}{d \varrho^k}(0) \neq 0,~~~\rho(\infty )
=\sqrt{\dfrac{2(\mu +\gamma )}{(\lambda +\alpha+\beta )}}, \nn\\
&f(0)=\pi,~~~~~f(\infty ) =0, \nn\\
&A(0)=-\dfrac{k}n,~~~~~A(\infty )=1,
\eea
and obtain a different D-type vortex whose magnetic
flux is given by
\bea
&\Phi =\Big(1+ \dfrac{k}{n} \Big) \dfrac{2\pi n}g=\dfrac{2\pi}{g}(n+k).
\eea
The $6\pi/g$-flux vortex with $n=1$
and $k=2$ is shown in Fig. \ref{sv6pi}. This tells that there exist
infinitely many D-type vortices which have the same
topology $\pi_2(S^2)=n$. Again this
is completely unexpected.

\subsection{$2\pi/g$-flux vortex}

\begin{figure}[t]
\includegraphics[scale=0.4]{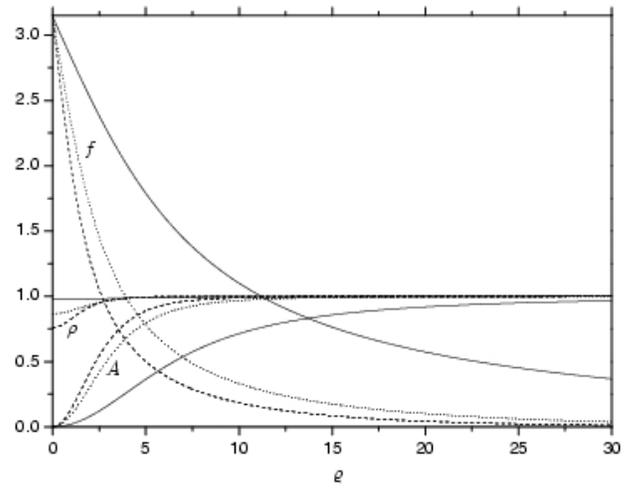}
\caption{The N-type straight vortex with $n=1$ and $2\pi/g$ flux.
Three solutions are shown: $\alpha=\beta=\gamma=0$ (solid
lines), $\alpha=\beta=0$, $\gamma =0.005$ (dashed lines),
and $\alpha=\gamma=0$, $\beta =-0.005$ (dotted lines).
Here the unit of the scale is $1/{\bar \rho}$ and
we have put $\lambda/g^2=2$.}
\label{sv2pi}
\end{figure}

Now we choose the Neumann boundary condition
at the core and the integer flux vacuum at the infinity \cite{ijpap,cm2},
\begin{eqnarray}
&\rho(0) \neq 0,~~~\dot \rho(0)=0,~~~\rho (\infty)
=\sqrt{\dfrac{2(\mu +\gamma )}{(\lambda +\alpha +\beta )}}, \nn\\
&f(0)=\pi,~~~~~f(\infty ) =0, \nn\\
&A(0)=0,~~~~~A(\infty )=1,
\end{eqnarray}
and find the vortex solutions. The solutions with $n=1$
but with different parameters are shown in Fig.~\ref{sv2pi}.
We call this a N-type vortex, because this comes from
the Neumann boundary condition at the core.
In this case $\phi_1$ behavior is the same as before.
But notice that $\phi_2$ has a maximum concentration at the core,
and approaches zero at the infinity.
This is a generic feature of a N-type vortex.

The magnetic flux of the vortex is given by
\begin{eqnarray}
&\Phi =\dfrac{}{} \int F_{\varrho \varphi}d^2x
= \dfrac{}{} \int \partial_\varrho A_\varphi d^2x \nn\\
&=\dfrac{2\pi n}g,
\end{eqnarray}
so that it has the same flux as the Abrikosov vortex.
But notice that the topology of the $CP^1$ field $\hn$ is
still non-Abelian as before, $\pi_2(S^2)=n$.
The reason why there exist two types of vortices which have
different magnetic fluxes but have the same topology
is that the magnetic flux is determined by
the boundary condition $A(\infty)-A(0)$, not by
the topology. The topology assures only the quantization of the flux,
and does not determine what is the unit flux quantum.

\subsection{Fractional flux vortex}

This is possible when we have the fractional flux vacuum
at infinity
\begin{eqnarray}
&\rho (\infty )=2\sqrt{\dfrac{2 \beta \mu -\alpha \gamma}
{4\beta \lambda-\alpha^2}},
~~~\cos f(\infty )=\dfrac{2\gamma \lambda -\alpha \mu}
{2\beta \mu-\alpha \gamma}, \nn\\
&A(\infty )=\dfrac 12 \dfrac{2(\gamma \lambda + \beta \mu)
-\alpha(\mu+\gamma)}{2\beta \mu-\alpha \gamma}.
\end{eqnarray}
At the core we can impose either the Dirichlet condition
(\ref{dbc}) or the Neumann condition (\ref{nbc}).

\begin{figure}[t]
\includegraphics[scale=0.4]{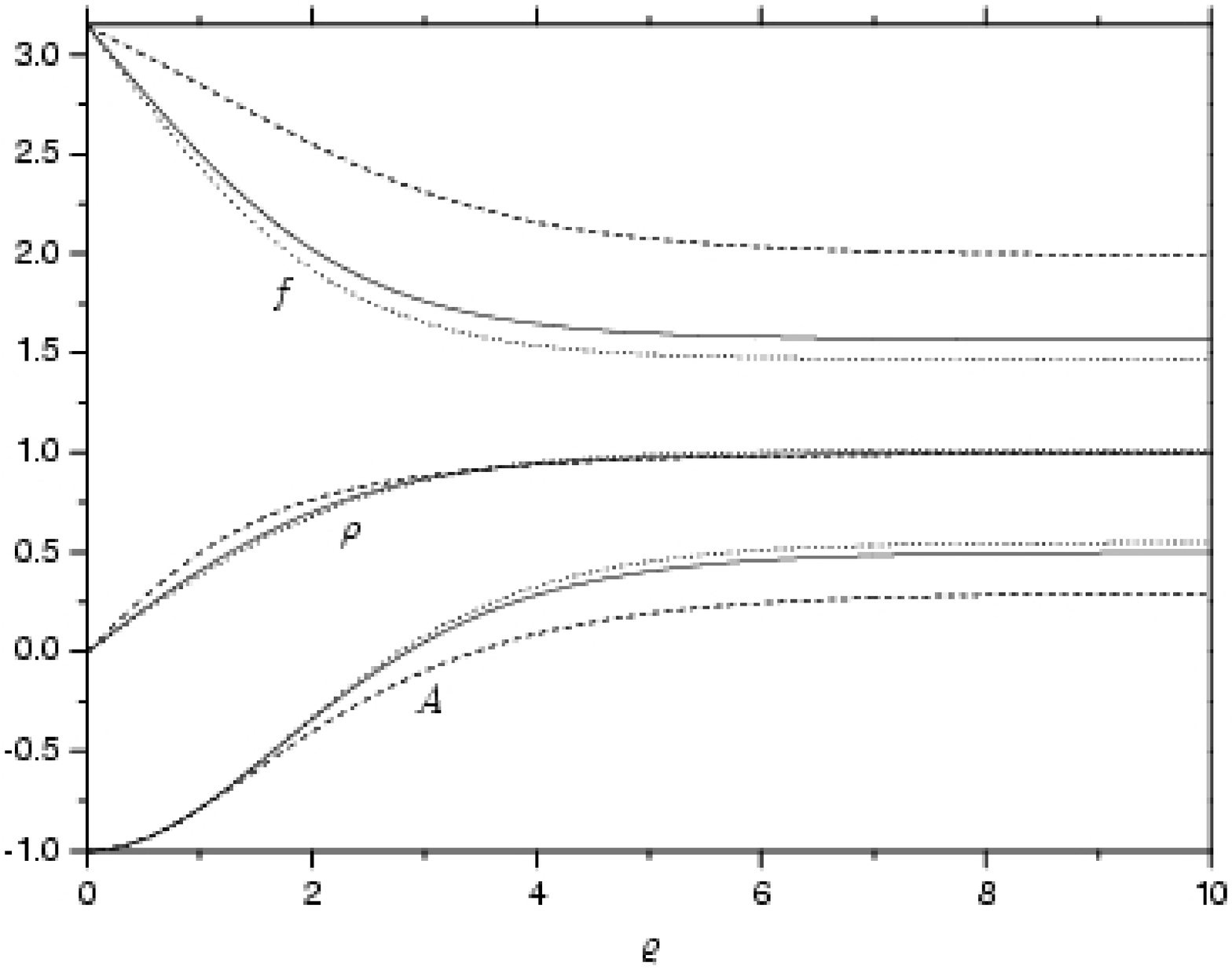}
\caption{The D-type straight vortices with $n=1$  and $k=1$
which have a fractional flux with
$\alpha=\gamma=0,~\beta=1.0$ (solid lines),
$\alpha=0,~\beta=\lambda,~\gamma=-0.2$ (dashed lines),
and $\alpha=-0.25,~\beta=\lambda,~\gamma=0$ (dotted lines).
Here the unit of the scale is $1/{\bar \rho}$ and
we have put $\lambda/g^2=2$.}
\label{sf4pi}
\end{figure}

We consider two special cases: \\
1. $\lambda _{11}=\lambda _{22}$ ($\alpha =0$).
In this case we have
\begin{eqnarray}
\rho(\infty )=\sqrt {\dfrac{2\mu}{\lambda}},
~~~~\cos f(\infty)=\dfrac{\gamma \lambda}{\beta \mu}.
\end{eqnarray}
So with the Dirichlet boundary condition at the core
the magnetic flux is given by
\begin{equation}
\Phi=\Big(\dfrac{\gamma \lambda}
{2\beta \mu} +\dfrac12+\dfrac{k}{n} \Big) \dfrac{2\pi n}{g},
\end{equation}
but with the Neumann boundary condition at the core
the magnetic flux is given by
\begin{equation}
\Phi=\Big(\dfrac{\gamma \lambda}{2\beta \mu} +\dfrac12 \Big)
\dfrac{2\pi n}{g}.
\end{equation}
Clearly they are fractional. \\
2. $\lambda_{12}=0$ ($\lambda=\beta$). In this case two
condensates $\phi_1$ and $\phi_2$ have no direct coupling,
and we have
\begin{equation}
<|\phi _1|> =\sqrt{\dfrac{\mu_1}{\lambda_{11}}},
~~~~~<|\phi _2|> =\sqrt{\dfrac{\mu _2}{\lambda_{22}}}.
\end{equation}
With this we have
\begin{eqnarray}
&\rho(\infty)=2 \sqrt {\dfrac{2\lambda \mu-\alpha\gamma}
{4\lambda ^2-\alpha^2}},
~~~\cos f(\infty )=\dfrac{2\gamma \lambda -\alpha \mu}
{2\lambda \mu-\alpha \gamma}, \nn\\
&A(\infty)=\dfrac 12 \dfrac{(2\lambda-\alpha)(\mu+\gamma)}
{2\mu \lambda -\alpha \gamma}.
\end{eqnarray}
So with the Dirichlet boundary condition at the core
the magnetic flux is given by
\begin{equation}
\Phi=\Big(\dfrac 12 \dfrac{(2\lambda-\alpha)(\mu+\gamma)}
{2\lambda \mu-\alpha \gamma}
+\dfrac{k}{n} \Big) \dfrac{2\pi n}{g},
\end{equation}
but with the Neumann boundary condition at the core
the magnetic flux is given by
\begin{equation}
\Phi=\dfrac 12 \dfrac{(2\lambda-\alpha)(\mu+\gamma)}
{2\lambda \mu-\alpha \gamma} \dfrac{2\pi n}{g}.
\end{equation}
Again they are fractional, in spite of the fact that
$\phi_1$ and $\phi_2$ have no direct coupling.
This is because they are coupled through the electromagnetic
potential, which tells that the two-gap superconductor is
not a naive superposition of two one-gap superconductor.

\begin{figure}[t]
\includegraphics[scale=0.4]{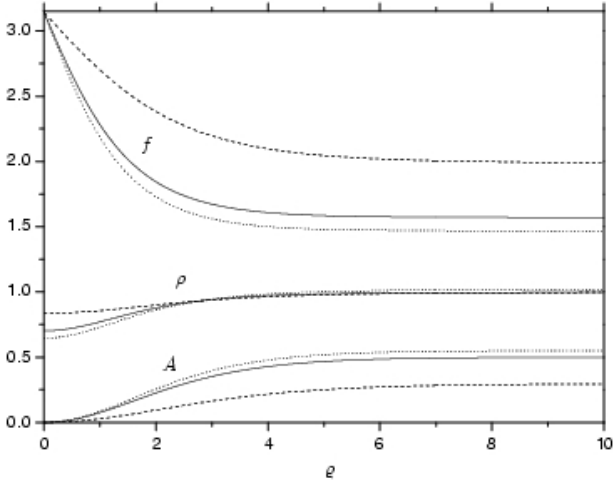}
\caption{The N-type straight vortices with $n=1$ which have
a fractional flux with $\alpha=\gamma=0,~\beta=1.0$ (solid lines),
$\alpha=0,~\beta=\lambda,~\gamma=-0.2$ (dashed lines),
and $\alpha=-0.25,~\beta=\lambda,~\gamma=0$ (dotted lines).
Here the unit of the scale is $1/{\bar \rho}$ and
we have put $\lambda/g^2=2$.}
\label{sf2pi}
\end{figure}

According to the different boundary condition
at the core there are two types of fractional flux vortices,
D-type and N-type. These fractional flux vortices with $n=1$
are plotted in Fig.~\ref{sf4pi} and Fig.~\ref{sf2pi}.
The fractional vortex is also topological,
but the topology of the fractional vortex is different from
that of integer flux vortex.
Notice that for the fractional flux vortices
the $\pi_2(S^2)$ topology of $\hn$ becomes trivial, $\pi_2(S^2)=0$.
This is because $\hn$ does not cover the target space
$S^2$ fully. But in this case we still have
a $U(1)$ topology $\pi_1(S^1)$, the topology of the $U(1)$
symmetry which leaves $\hn$ invariant. And this Abelian
topology describes the topology of
the fractional flux vortex. So the topology of
the fractional flux vortices is the same as that of
the Abrikosov vortex.

An important feature of the fractional flux vortex is that
the energy per unit length of the vortex is logarithmically
divergent, which can be shown from the Hamiltonian (\ref{scfe4}).
This is because the fractional flux vortex has a non-vanishing
neutral current $k_\mu$ at the infinity.
This, however, does not make the fractional flux vortex unphysical.
In laboratory setting one can observe such vortex because
one has a natural cutoff
parameter $\Lambda$ fixed by the size of the superconductor,
which can effectively make the
energy of the fractional flux vortex finite.
Indeed in $\rm He^3$ superfluid one often encounters
the vorticity vortex whose energy is logarithmically
divergent \cite{pra05,vol}.

The existence of fractional flux vortices
in two-gap superconductor has been pointed out
before in London limit \cite{baba2}. Our analysis in this paper
shows that the London limit does not fully describe the vortex
in two-gap superconductor. This is because the magnetic flux
is determined by the boundary condition at the origin
(as well as the boundary condition at the infinity).
Clearly the existence of two types of vortices
which have different core structure and
different magnetic flux can not be understood in London limit.

In this section we have shown that the two-gap superconductor
can have totally different magnetic vortices which can not be
found in ordinary superconductor. There are two types of vortices,
D-type and N-type, and both have two different topologies, $\pi_2(S^2)$
and $\pi_1(S^1)$, which describe the vortices.
The integral flux vortex is described by the $\pi_2(S^2)$
topology, but the fractional vortex is described by the
$\pi_1(S^1)$ topology. As importantly, there are infinitely
many different vortices within the same topological sector.
Moreover, the magnetic flux
of the D-type vortex is larger than
that of the N-type vortex by a factor $2\pi k/g$,
so that in the same topological sector the D-type vortex
has more energy than the N-type vortex.

Obviously all these vortices are non-Abrikosov. This does
not mean that two-gap superconductor can not admit an
Abrikosov vortex. With $f=\pi$ (or $f=0$) and
$\alpha=\beta=\gamma=0$, (\ref{sveq}) describes an
Abrikosov vortex. This is because with $\phi_1=0$
(or with $\phi_2=0$) the two-gap superconductor reduces to an
ordinary superconductor.

\section{Helical vortex}

In this section we show that the above non-Abrikosov
vortices can be twisted to form a twisted magnetic vortex.
With the twisting we obtain the helical vortex
which is periodic in $z$-coordinate. To show this
we choose the following ansatz \cite{ijpap,cm2},
\begin{eqnarray}
&\rho =\rho (\varrho ),~~~~~\xi =\left(
\begin{array}{c}
\cos \dfrac{f(\varrho )}2\exp (-in\varphi )\nonumber \\
\sin \dfrac{f(\varrho )}2\exp (imkz)
\end{array}
\right) ,  \nonumber \\
&A_\mu =\dfrac 1g\Big(nA_1(\varrho )\partial _\mu \varphi
+mkA_2(\varrho )\partial _\mu z\Big).
\label{hvans}
\end{eqnarray}
Obviously the ansatz is periodic in $z$-coordinate,
with the period $2\pi/k$.

With the ansatz we have
\bea
&\hat{n}=\xi ^{\dag }\vec{\sigma}\xi =\left(
\begin{array}{c}
\sin f(\varrho )\cos (n\varphi +mkz) \\
\sin f(\varrho )\sin (n\varphi +mkz) \\
\cos f(\varrho ) \end{array} \right),  \nonumber \\
&C_\mu =n\dfrac{\cos {f(\varrho )}+1}g \partial _\mu \varphi
+mk \dfrac{\cos {f(\varrho )}-1}g \partial _\mu z, \nn\\
&j_\mu =g\rho ^2\Big(n\big(A_1-\dfrac{\cos {f}+1}2\big)
\partial _\mu \varphi   \nonumber \\
&+mk\big(A_2-\dfrac{\cos {f}-1}2\big) \partial _\mu z\Big), \nn\\
& k_\mu = g\rho ^2\Big(n\big(A_1 \cos f-\dfrac{\cos {f}+1}2\big)
\partial _\mu \varphi   \nonumber \\
&+mk\big(A_2 \cos f+\dfrac{\cos {f}-1}2\big) \partial _\mu z\Big),
\label{schv}
\eea
and the following Hamiltonian
\bea
&{\cal H} =\dfrac 12\dot{\rho}^2+\dfrac 18\rho^2
\Big(\dot{f}^2+ \big( \dfrac{n^2}{\varrho^2}
+m^2 k^2 \big ) \sin ^2f \Big)  \nn\\
&+ \dfrac{\rho^2}2 \Big[\dfrac{n^2}{\varrho^2}
\Big( A_1 - \dfrac{\cos f+1}{2}\Big)^2
+ m^2k^2\Big(A_2- \dfrac{\cos f -1}{2} \Big)^2 \Big] \nn\\
&+\dfrac {1}{2g^2}\Big( \dfrac{n^2}{\varrho^2}\dot{A_1}^2+m^2k^2
\dot{A_2}^2 \Big)
+\dfrac \lambda 8  \Big[(\rho ^2-\bar \rho^2)^2  \nn\\
&+\dfrac{\alpha}\lambda (\rho^2-\dfrac{4\gamma }\alpha)\rho^2 \cos f
+\dfrac{\beta}\lambda \rho^4\cos ^2f \Big]-\dfrac{\mu^2}{2\lambda}.
\label{hvham}
\eea
With this (\ref{sceq2}) becomes
\begin{eqnarray}
&\ddot{\rho}+\dfrac 1\varrho \dot{\rho}-\Big[\dfrac 14
\big(\dot{f}^2+(\dfrac{n^2}{\varrho^2}+m^2k^2)\sin^2f \big)  \nonumber \\
&+\dfrac{n^2}{\varrho^2}(A_1-\dfrac{\cos {f}+1}2 )^2
+m^2k^2(A_2- \dfrac{\cos {f}-1}2)^2\Big]\rho   \nonumber \\
&=\dfrac \lambda 2\Big[\big(\rho ^2-\rho _0^2\big) +
\dfrac{\alpha}{\lambda} \big(\rho^2-\dfrac{2\gamma}\alpha \big)
\cos f +\dfrac{\beta}{\lambda} \rho^2\cos^2 f \Big] \rho,  \nonumber \\
&\ddot{f}+\big(\dfrac 1\varrho +2\dfrac{\dot{\rho}}\rho %
\big)\dot{f}-2\Big[\dfrac{n^2}{\varrho^2}\big(A_1-
\dfrac 12\big)  \nonumber \\
&+m^2k^2\big(A_2+\dfrac 12\big) \Big] \sin f \nonumber \\
&=\Big(2\gamma - \big(\dfrac \alpha 2
+\beta \cos f \big) \rho^2 \Big) \sin f,  \nonumber \\
&\ddot{A_1}-\dfrac 1\varrho \dot{A}_1-g^2\rho ^2\Big(A_1
- \dfrac{\cos {f}+1}2 \Big)=0,  \nonumber \\
&\ddot{A_2}+\dfrac 1\varrho \dot{A}_2-g^2\rho ^2\Big(A_2
-\dfrac{\cos {f}-1}2 \Big)=0.
\label{hveq}
\end{eqnarray}
This is an obvious generalization of (\ref{sveq}).

\begin{figure}[t]
\includegraphics[scale=0.4]{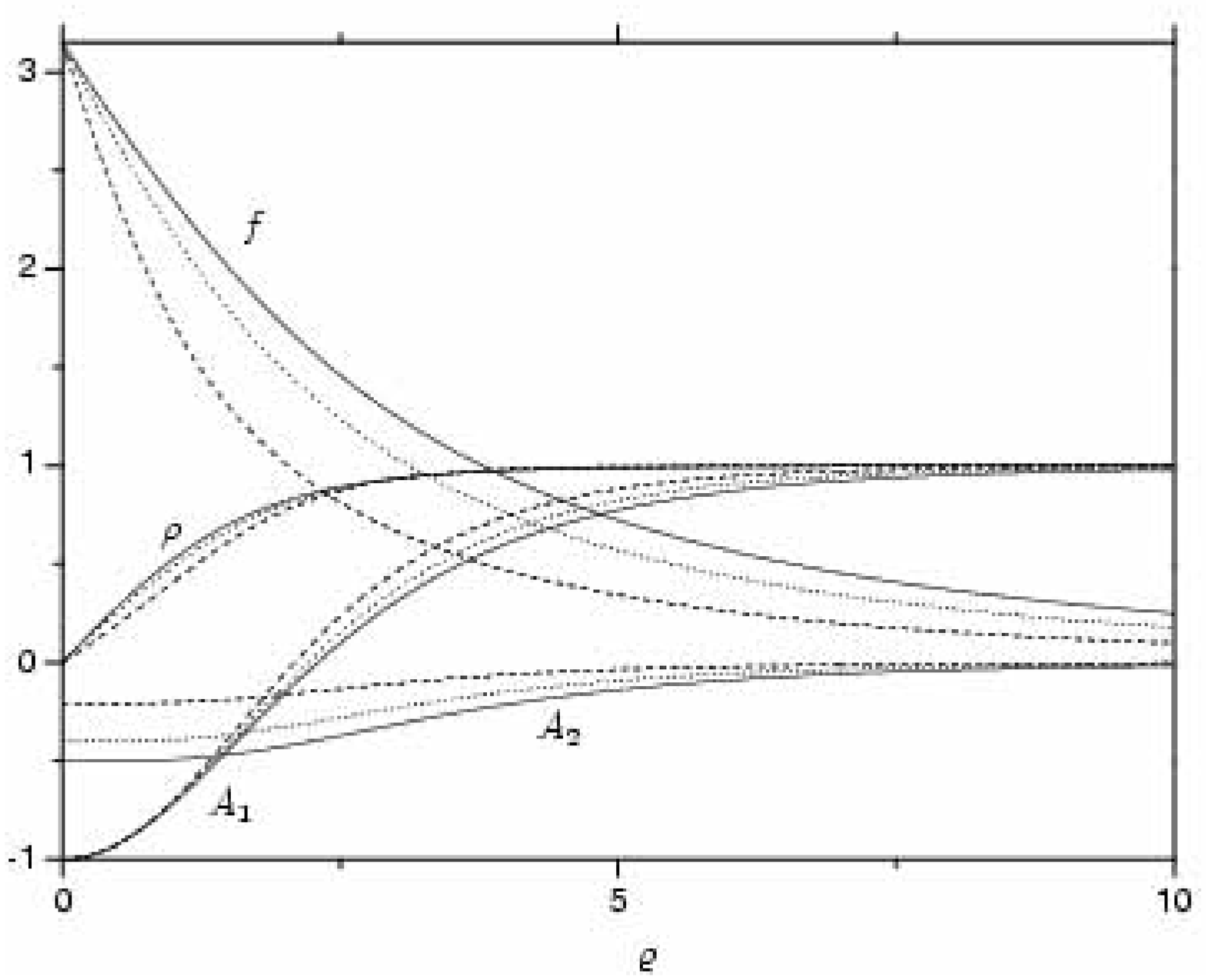}
\caption{The D-type helical vortex with $4\pi/g$-flux along the vortex
with $m=n=1$ (with $k=1$). Three solutions are shown:
$\alpha=\beta=\gamma=0$ (solid lines),
$\alpha=\beta=0$ and $\gamma =0.005$ (dashed lines),
$\alpha=\gamma=0$ and $\beta =-0.005$ (dotted lines).
Here the unit of the scale is $1/{\bar \rho}$ and
we have put $k=0.12 \bar \rho$ and $\lambda/g^2=2$.}
\label{hv4pi}
\end{figure}

\begin{figure}[t]
\includegraphics[scale=0.4]{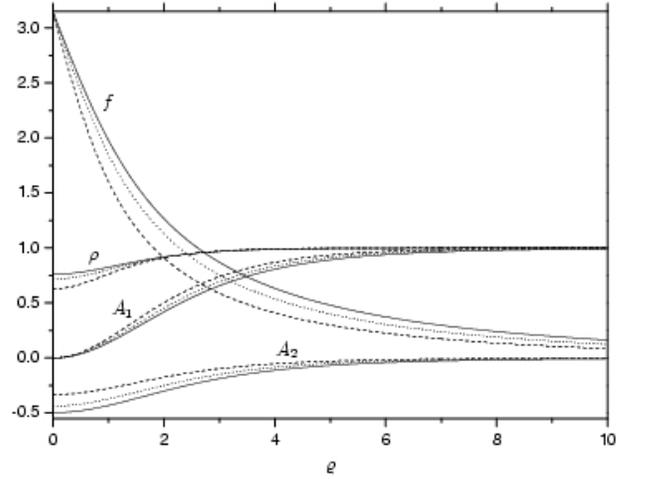}
\caption{The N-type helical vortex with $2\pi/g$-flux along the vortex
with $m=n=1$. Three solutions are shown:
$\alpha=\beta=\gamma=0$ (solid lines),
$\alpha=\beta=0$ and $\gamma =0.005$ (dashed lines),
$\alpha=\gamma=0$ and $\beta =-0.005$ (dotted lines).
Here the unit of the scale is $1/{\bar \rho}$ and
we have put $k=0.12 \bar \rho$ and $\lambda/g^2=2$.}
\label{hv2pi}
\end{figure}

To obtain the helical vortex we first consider the integer flux
boundary condition at the infinity
\begin{eqnarray}
&\rho (\infty )=\sqrt {\dfrac{2(\mu +\gamma )}
{(\lambda +\beta +\alpha)}},~~~~~f(\infty )=0, \nn\\
&A_1(\infty )=1,~~~~~A_2(\infty )=0.
\end{eqnarray}
Just like the straight vortex, there are two types of
boundary conditions at the core. Here we consider only the case $n=1$
for simplicity: \\
A. Dirichlet boundary condition
\bea
&\rho(0)=0,~~~~~f(0)=\pi, \nn\\
&A_1(0)=-1,~~~~~\dot A_2(0)=0.
\label{dbchv}
\eea
B. Neumann boundary condition
\bea
&\dot{\rho}(0)=0,~~~~~f(0)=\pi, \nn\\
&A_1(0)=0,~~~~~\dot A_2(0)=0.
\label{nbchv}
\eea
With the Dirichlet boundary condition we have the
D-type helical vortex which has $4\pi/g$-flux along the
vortex shown in Fig.~\ref{hv4pi}, but with the Neumann boundary
condition we have the N-type helical vortex which has
$2\pi/g$-flux along the vortex shown in Fig.~\ref{hv2pi}.

\begin{figure}[t]
\includegraphics[scale=0.4]{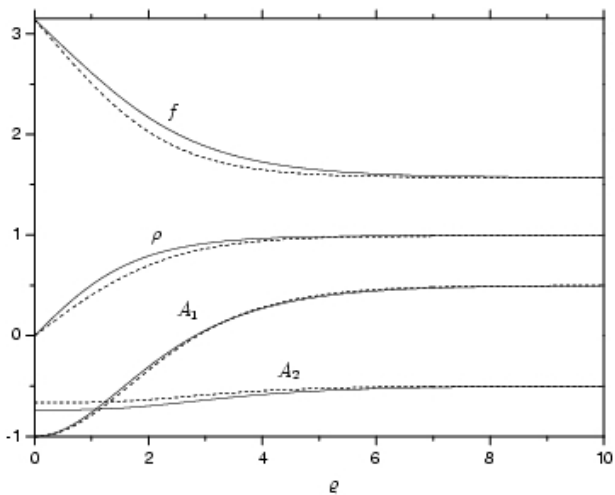}
\caption{The D-type helical vortices which have a fractional flux
with $m=n=1$ (with $k=1$). Two solutions with $\alpha=\gamma=0,~\beta= 0.5$
(solid lines), and $\alpha=\gamma=0,~\beta = 1.0$ (dashed lines)
are shown. Here the unit of the scale is $1/{\bar \rho}$ and
we have put $\lambda/g^2=2,~k=0.1 \bar \rho$.}
\label{hf4pi}
\end{figure}

\begin{figure}[t]
\includegraphics[scale=0.4]{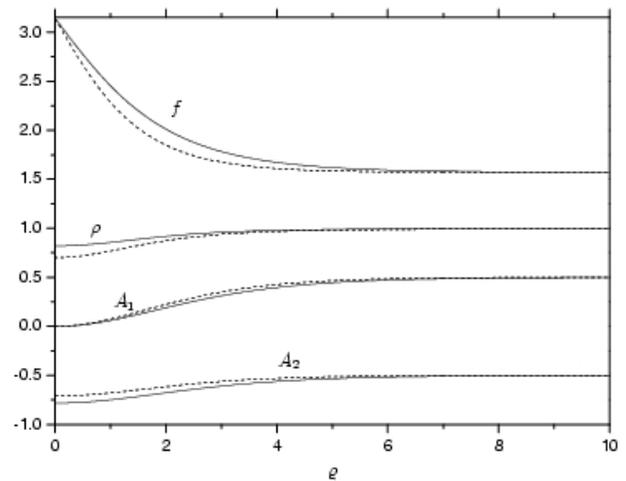}
\caption{The N-type helical vortices which have a fractional flux
with $m=n=1$. Two solutions with $\alpha=\gamma=0,~\beta= 0.5$
(solid lines), and $\alpha=\gamma=0,~\beta = 1.0$ (dashed lines)
are shown. Here the unit of the scale is $1/{\bar \rho}$ and
we have put $\lambda/g^2=2,~k=0.1 \bar \rho$.}
\label{hf2pi}
\end{figure}

For the fractional flux helical vortex we impose the
fractional flux boundary condition at the infinity
\begin{eqnarray*}
&\rho (\infty )=2\sqrt{\dfrac{2\beta \mu-\alpha \gamma}
{4\beta \lambda-\alpha ^2}},
~~~\cos f(\infty )=\dfrac{2\gamma \lambda -\alpha \mu}
{2\beta \mu -\alpha \gamma}, \nn\\
&A_1(\infty )=\dfrac 12 \dfrac{2(\gamma \lambda+\beta \mu)
-\alpha(\mu+\gamma)}{2\beta \mu -\alpha \gamma}, \nn\\
&A_2(\infty )=\dfrac 12 \dfrac{2(\gamma \lambda-\beta \mu)
-\alpha(\mu-\gamma)}{2\beta \mu -\alpha \gamma}.
\end{eqnarray*}
Now, with the Dirichlet boundary condition at the core,
we obtain the D-type helical vortex which has a fractional flux along
the vortex shown in Fig.~\ref{hf4pi}.
But with the Neumann boundary
condition (\ref{nbchv}) at the core,
we obtain the N-type helical vortex which has a fractional flux along
the vortex shown in Fig.~\ref{hf2pi}.
Notice that, just as the fractional flux straight vortex,
the energy per one period of the fractional flux helical vortex
is logarithmically divergent.

A new feature of the helical vortex is that the magnetic
flux becomes helical.
Indeed the ansatz (\ref{hvans}) tells that the magnetic
flux can be decomposed to the one along
the vortex and the other around the vortex \cite{ijpap,cm2}
\bea
&F_{\hat \varrho \hat \varphi}=\dfrac ng\dfrac{\dot{A}_1}\varrho,
~~~F_{\hat{z}\hat{\varrho}}=-\dfrac{mk}g\dot{A}_2,
\eea
so that we have two magnetic fluxes linked together,
\bea
&\Phi_{\hat z}=\dfrac{}{}\int F_{\hat{\varrho}\hat{\varphi}}\varrho
d\varrho d\varphi =\Big(A_1(\infty )-A_1(0)\Big)\dfrac{2\pi n}g, \nn\\
&\Phi_{\hat \varphi}=\dfrac{}{}\int F_{\hat{z}\hat{\varrho}}dzd\varrho
=-\Big(A_2(\infty )-A_2(0)\Big)\dfrac{2\pi m}g.
\eea
Obviously $\Phi_{\hat \varphi}$ is due to the helical structure
of the vortex, which becomes fractional in general.

Another important feature of the helical vortex is that
the electromagnetic current $j_\mu$ which is responsible for
the Meissner effect also becomes helical.
In particular, it has a non-trivial electromagnetic current
$j_{\hat z}$ along the vortex which generates the magnetic flux
$\Phi_{\hat \varphi}$, in addition to the usual electromagnetic
current $j_{\hat \varphi}$ around the vortex which is responsible
for $\Phi_{\hat z}$. But notice that the total electromagnetic
current $i_{\hat z}$ along the vortex becomes zero.
This, together with (\ref{sc}), tells that $\phi_1$ and $\phi_2$
generate non-vanishing electromagnetic currents $i_{\hat z}^{(1)}$
and $i_{\hat z}^{(2)}$ which flow oppositely and cancel each other.
In this sense we may call the helical vortex superconducting,
even though it has no net electromagnetic current
$i_{\hat z}$ along the vortex \cite{ijpap,cm2}.

\section{magnetic knot in two-gap superconductor}

Clearly the helical vortex is unstable unless the periodicity
condition is enforced by hand. Nevertheless it has an important
implication, because the helical vortex predicts the existence of a
topological knot in two-gap superconductor. This is because
we can make it a twisted magnetic vortex ring
smoothly bending and connecting two
periodic ends together. The resulting twisted magnetic vortex ring
becomes a knot whose topology is described by the Chern-Simon index
of the electromagnetic potential \cite{ijpap,cm2}.

There have been two objections against the existence of
a stable magnetic vortex ring in Abelian superconductor.
First, it is supposed to be unstable
due to the tension created by the ring \cite{huang}.
Indeed if one constructs a vortex ring from an Abrikosov vortex,
it becomes unstable because of the tension.
But we can easily overcome this difficulty by twisting
the magnetic vortex first and connecting the periodic ends together.
In this case the non-trivial twist of the magnetic field
forbids the untwisting of the vortex ring
by any smooth deformation of field configuration, and
the vortex ring becomes a stable knot.
The other objection is that the Abelian gauge theory
is supposed to have no non-trivial knot topology
which allows a stable vortex ring.
This again is a common misconception. As we have seen,
the theory has a well-defined knot
topology $\pi_3(S^2)$ described by the Chern-Simon index
of the electromagnetic potential.
This tells that there is
no reason whatsoever why the Abelian superconductor can not have
a topological knot.

To demonstrate the existence of a topological knot in the two-gap
superconductor, we introduce the toroidal coordinates
($\eta,\gamma,\varphi $) defined by
\begin{eqnarray}
&x=\dfrac{a}{D}\sinh{\eta}\cos{\varphi}, ~~~y=\dfrac{%
a}{D}\sinh{\eta}\sin{\varphi},  \nonumber \\
&z=\dfrac{a}{D}\sin{\gamma},  \nonumber \\
&D=\cosh{\eta}-\cos{\gamma},  \nonumber \\
&ds^2=\dfrac{a^2}{D^2} \Big(d\eta^2+d\gamma^2+\sinh^2\eta
d\varphi^2 \Big),  \nonumber \\
&d^3x=\dfrac{a^3}{D^3} \sinh{\eta} d\eta d\gamma d\varphi,
\label{tc}
\end{eqnarray}
where $a$ is the radius of the knot defined by $\eta=\infty$.
Notice that in toroidal coordinates, $\eta=\gamma=0$ represents
spatial infinity of $R^3$, and $\eta=\infty$ describes the torus
center.

Now we choose the following ansatz,
\begin{eqnarray}
&\phi=\dfrac{1}{\sqrt 2} \rho (\eta, \gamma) \Bigg(\matrix{\cos
\dfrac{f(\eta,\gamma )}{2} \exp (-im\varphi ) \cr \sin
\dfrac{f(\eta ,\gamma)}{2} \exp (in\omega (\eta,\gamma))} \Bigg),
\nonumber \\
&A_\mu =\dfrac ng A_0(\eta, \gamma) \partial_\mu \eta
+\dfrac ng A_1(\eta, \gamma) \partial_\mu \gamma \nn\\
&+\dfrac mg A_2(\eta, \gamma)\partial_\mu \varphi.
\label{sckans}
\end{eqnarray}
With this we have
\begin{eqnarray}
&\hn=\Bigg(\matrix {
\sin f(\eta,\gamma )\cos (n\omega +m\varphi) \cr
\sin f(\eta,\gamma )\sin (n\omega +m\varphi) \cr
\cos f(\eta,\gamma ) } \Bigg), \nn\\
&C_\mu = \dfrac{m}{2g} (\cos f +1)\partial_\mu \varphi +
\dfrac{n}{2g} (\cos f -1) \partial_\mu \omega, \nn\\
&F_{\eta \gamma }=\dfrac ng \big (\partial_\eta A_1
-\partial_\gamma A_0 \big), \nn\\
&F_{\gamma\varphi }=\dfrac mg \partial_\gamma A_2,
~~~~~F_{\varphi \eta }=-\dfrac mg \partial_\eta A_2.
\label{kvf1}
\end{eqnarray}
Notice that, in the orthonormal frame $(\hat \eta, \hat \gamma,
\hat \varphi)$, we have
\begin{eqnarray}
&A_{\hat{\eta}}=\dfrac Da A_\eta,
~~A_{\hat{\gamma}}=\dfrac Da A_\gamma,
~~A_{\hat{\varphi}}=\dfrac {D}{a \sinh \eta} A_\varphi, \nn\\
&F_{\hat{\eta}\hat{\gamma}}=\dfrac {nD^2}{g a^2}
\big (\partial_\eta A_1-\partial_\gamma A_0 \big), \nn\\
&F_{\hat{\gamma}\hat{\varphi}}=\dfrac {mD^2}{g a^2 \sinh \eta}
\partial_\gamma A_2, \nn\\
&F_{\hat{\varphi}\hat{\eta}}=-\dfrac {mD^2}{ga^2 \sinh \eta}
\partial_\eta A_2. \nn
\label{kvf2}
\end{eqnarray}
Next, we adopt the $SU(2)$ symmetric potential
with $\lambda_{11}=\lambda_{22}=\lambda_{12}=\lambda$ and
$\mu_1=\mu_2=\mu$ for simplicity.
In this case we have the following knot equation
\begin{widetext}
\begin{eqnarray}
&\Bigg[ \partial _\eta ^2+\partial _\gamma ^2
+\Big( \dfrac{\cosh \eta }{\sinh \eta }-\dfrac{\sinh \eta }D\Big)
\partial _\eta -\dfrac{\sin \gamma }D\partial _\gamma \Bigg] \rho
-\dfrac 14\Bigg[  \big(\partial _\eta f \big)^2
+ \big(\partial_\gamma f \big)^2 \nn\\
&+\sin ^2f\Big( n^2 \big(\partial_\eta \omega
 \big)^2+n^2 \big(\partial_\gamma \omega
\big)^2+\dfrac{m^2}{\sinh ^2\eta }\Big) \Bigg] \rho
-\Bigg[ n^2\Big(A_0- \dfrac{\cos f +1}2 \partial_\eta \omega \Big)^2
+n^2 \Big(A_1- \dfrac{\cos f +1}2 \partial_\gamma \omega \Big)^2 \nn\\
&+\dfrac{m^2}{\sinh ^2\eta } \Big(A_2
- \dfrac{\cos f -1}2 \Big)^2\Bigg] \rho
=\dfrac \lambda 2\dfrac{a^2}{D^2}
\Big(\rho ^2-\bar \rho^2 \Big)\rho, \nn\\
&\Bigg[ \partial _\eta ^2+\partial_\gamma ^2
+\Big( \dfrac{\cosh \eta }{\sinh \eta }-\dfrac{\sinh \eta }D\Big)
\partial _\eta -\dfrac{\sin \gamma }D
\partial _\gamma \Bigg] f+\dfrac 2\rho \Big(\partial _\eta
f\partial _\eta \rho
+\partial _\gamma f\partial _\gamma \rho  \Big) \nn\\
&=2\sin f\Bigg[ n^2 \Big(A_0-\dfrac 12\partial _\eta \omega
 \Big)\partial _\eta \omega +n^2 \Big(A_1
-\dfrac 12\partial _\gamma \omega \Big)\partial _\gamma \omega
+\dfrac{m^2}{\sinh ^2\eta } \Big(A_2+\dfrac 12 \Big)\Bigg], \nn\\
&\Bigg[ \partial _\eta ^2+\partial _\gamma ^2+\Big( \dfrac{\cosh \eta }{%
\sinh \eta }-\dfrac{\sinh \eta }D\Big) \partial _\eta -\dfrac{\sin \gamma }%
D\partial _\gamma \Bigg] \omega +\dfrac 2\rho \Big(\partial _\eta
\rho \partial_\eta \omega +\partial _\gamma \rho
\partial_\gamma \omega \Big)
-\dfrac{\sin f}{1+\cos f} \Big(\partial _\eta f\partial _\eta
\omega +\partial_\gamma f\partial_\gamma \omega \Big) \nn\\
&+\dfrac2{\sin f} \Big(\partial_\eta fA_0+\partial_\gamma fA_1 \Big)=0, \nn\\
& \Big(\partial _\gamma +\dfrac{\sin \gamma }D \Big)\Big(\partial
_\gamma A_0-\partial _\eta A_1 \Big)=\dfrac{a^2}{D^2}g^2\rho ^2
\Big(A_0- \dfrac{\cos f +1}2 \partial_\eta \omega \Big), \nn\\
& \Big(\partial _\eta +\dfrac{\cosh \eta }{\sinh \eta }
+\dfrac{\sinh \eta }D \Big) \Big(\partial _\eta A_1-\partial _\gamma A_0
\Big)=\dfrac{a^2}{D^2}g^2\rho ^2 \Big(A_1
- \dfrac{\cos f +1}2 \partial_\gamma \omega \Big), \nn\\
&\Bigg[ \partial _\eta ^2+\partial _\gamma ^2-\Big( \dfrac{\cosh
\eta }{\sinh \eta }-\dfrac{\sinh \eta }D\Big)
\partial _\eta +\dfrac{\sin \gamma }D
\partial _\gamma \Bigg] A_2=\dfrac{a^2}{D^2}g^2\rho ^2
\Big(A_2- \dfrac{\cos f -1}2  \Big).
\label{sckeq}
\end{eqnarray}
Moreover, from (\ref{scfe2}) and (\ref{sckans}) we have the following
Hamiltonian for the knot
\begin{eqnarray}
&{\cal H}=\dfrac{D^2}{2a^2}\Bigg[ (\partial _\eta \rho
)^2+(\partial _\gamma \rho )^2+\dfrac 14\rho ^2\big( (\partial
_\eta f)^2+(\partial _\gamma f)^2\big) +\dfrac 14\rho ^2\sin
^2f\Big( n^2(\partial _\eta \omega )^2+n^2(\partial
_\gamma \omega )^2+\dfrac{m^2}{\sinh ^2\eta }\Big) \Bigg] \nn\\
&+\dfrac{D^2}{2a^2}\Bigg[ n^2(A_0- \dfrac{\cos f +1}2
\partial_\eta \omega )^2+n^2(A_1-\dfrac{\cos f +1}2
\partial_\gamma \omega )^2+\dfrac{m^2}{\sinh
^2\eta }(A_2 -\dfrac{\cos f -1}2 )^2\Bigg] \rho ^2 \nn\\
&+\dfrac{D^4}{2g^2a^4}\Bigg[ n^2(\partial _\eta A_1-\partial _\gamma A_0)^2+%
\dfrac{m^2}{\sinh ^2\eta }\Big( (\partial _\eta A_2)^2+(\partial
_\gamma A_2)^2\Big) \Bigg]  +\dfrac \lambda 8(\rho ^2-\bar \rho
^2)^2.
\label{sckham}
\end{eqnarray}
\end{widetext}
Minimizing the energy we reproduce the knot equation
(\ref{sckeq}).

\begin{figure}[t]
\includegraphics[scale=0.4]{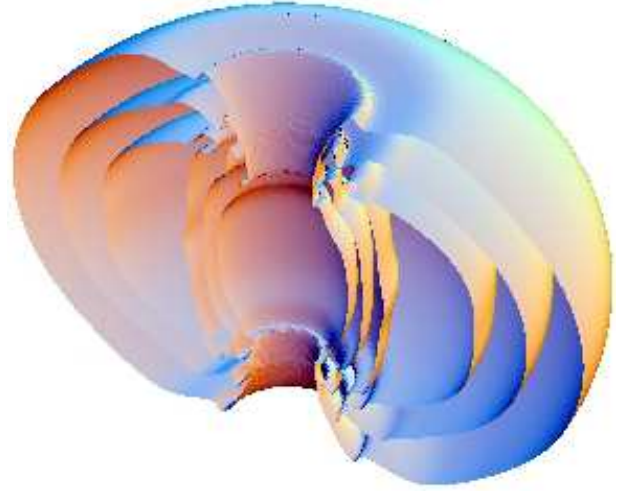}
\caption{The energy profile of a N-type knot with $m=n=1$. Here we have
put $\lambda/g^2=2$.} \label{2sck3d}
\end{figure}

In toroidal coordinates, $\eta=\gamma=0$ represents
spatial infinity of $R^3$, and $\eta=\infty$ describes
the torus center. So we can impose
the following Neumann boundary condition
\begin{eqnarray}
&\rho(0,0)=\bar \rho, ~~~~~\dot \rho(\infty,\gamma)=0,  \nonumber \\
&f(0,\gamma)=0, ~~~~~f(\infty,\gamma)=\pi,  \nonumber \\
&\omega(\eta,0)=0, ~~~~~\omega(\eta,2 \pi)=2 \pi,  \nonumber \\
&A_0(0,\gamma)=0, ~~~~~A_0(\infty,\gamma)=0,  \nonumber \\
&A_1(0,\gamma)=1, ~~~~~A_1(\infty,\gamma)=0,  \nonumber \\
&A_2(0,\gamma)=0, ~~~~~A_2(\infty,\gamma)=-1,
\label{sckbc}
\end{eqnarray}
to obtain the desired knot.
A numerical integration of (\ref{sckeq}) with the boundary
conditions (\ref{sckbc}) is difficult to perform.
But we can obtain the actual knot profile of $\rho,~f$, and
$\omega$ by minimizing the energy. From (\ref{sckham}) the knot
energy is given by
\begin{eqnarray}
&E=\dfrac{}{} \int {\cal H} \dfrac{a^3}{D^3} \sinh \eta d\eta
d\gamma d\varphi.
\label{scke1}
\end{eqnarray}
We find that, for $m=n=1$,
the radius of knot which minimizes the energy (\ref{scke1})
is given by
\begin{eqnarray}
a \simeq \dfrac{1.2}{\sqrt{2\mu}}.
\label{sckrad}
\end{eqnarray}
From this we obtain the three-dimensional energy profile
of the lightest axially
symmetric knot in two-gap superconductor, which is shown in
Fig.~\ref{2sck3d}.

With this we can estimate the energy of the axially symmetric
knot,
\begin{eqnarray}
E \simeq 51 \dfrac{\bar \rho}{\sqrt \lambda}.
\label{scke2}
\end{eqnarray}
We can also calculate the magnetic flux of the knot. Since the flux is
helical, we have two fluxes, the flux $\Phi_{\hat \gamma}$ passing
through the knot disk of radius $a$ in the $xy$-plane and the flux
$\Phi_{\hat \varphi}$ which surrounds it. From the knot solution
we find
\begin{eqnarray}
&\Phi _{\hat \eta} =\dfrac{}{}\int F_{\gamma \varphi }d\gamma d\varphi =0.
\nonumber\\
&\Phi _{\hat \gamma} =\dfrac{}{}\int F_{\varphi \eta }d\varphi d\eta
=\dfrac{2m\pi }g, \nn\\
&\Phi _{\hat \varphi} =\dfrac{}{}\int F_{\eta \gamma }d\eta d\gamma
=-\dfrac{2n\pi }g
\label{kflux}
\end{eqnarray}
The flux is quantized in the unit of $2\pi/g$, but this is
due to the $SU(2)$ symmetric potential and the Neumann
boundary condition (\ref{sckbc}). In general they can be fractional.
But independent of this the two fluxes are linked,
whose linking number becomes $mn$. This is important.

This confirms that the knot can be viewed as a twisted
magnetic vortex ring, where the linking of two magnetic fluxes
provides the knot topology. There is a natural
candidate which can describe this topology of twisted
magnetic vortex ring, the Chern-Simon index of
the electromagnetic potential which describes
the $\pi_3(S^2)$ topology.
We can calculate the knot quantum number
from the Chern-Simon index, and find
\begin{eqnarray}
Q_{CS} =-\dfrac{mn}{8\pi ^2}\int \varepsilon^{ijk} A_i
F_{jk}dx^3 = mn.
\label{sckqncs}
\end{eqnarray}
This confirms that the Chern-Simon index is indeed given by the
linking number of two magnetic fluxes $\Phi_{\hat \gamma}$
and $\Phi_{\hat \varphi}$.

It has been asserted that the knot topology is described by
the $\pi_3(S^2)$ topology of the $CP^1$ field $\hn$ \cite{baba1}.
We emphasize that this is only partially true,
which becomes correct only when the knot carries an integer
magnetic flux. Indeed, in this case
$\hn$ acquires a non-trivial knot topology
$\pi_3(S^2)$, which is given by \cite{ijpap}
\begin{eqnarray}
&Q =-\dfrac{mn}{8\pi ^2}\int (\partial_\eta f\partial_\gamma \omega
-\partial_\gamma f \partial_\eta \omega)
\sin f d\eta d\gamma d\varphi \nonumber \\
&= - \dfrac{mn}{4\pi} \int \sin f df d\omega  = mn,
\label{sckqn}
\end{eqnarray}
where the last equality comes from the boundary condition (\ref{sckbc}).
Notice, however, that the $\pi_3(S^2)$ topology of $\hn$
becomes trivial when the knot has a fractional flux. This is
because the fractional flux knot comes from a fractional flux vortex,
which has a trivial $\pi_2(S^2)$ topology. This means that
the $\pi_3(S^2)$ topology of $\hn$ can not describe the
topology of a fractional flux knot. In contrast,
the Chern-Simon index of the electromagnetic potential
can still provide the non-trivial $\pi_3(S^2)$,
even when the knot has a fractional flux.
This is because the Chern-Simon index describes
the knot topology of the twisted magnetic flux.

\section{Josephson Interaction}

It has been well-known that two-gap superconductor may allow
the prototype Josephson interaction \cite{maz}
\bea
\eta (\phi_1^{*}\phi_2 +\phi_1\phi_2^{*}),
\eea
where $\eta$ is a coupling constant. But we can consider
a more general quartic Josephson interaction,
\bea
&V_J = \bar \eta \Big[ \big(\phi_1^{*}\phi_2 \exp(i\theta_1)
+\phi_2^{*}\phi_1 \exp(-i\theta_1) \big) \Big]\nn\\
&+\bar \eta_1 \Big[(\phi_1^{*}\phi_2)^2 \exp(i\theta_2)
+(\phi_2^{*}\phi_1)^2 \exp(-i\theta_2) \Big] \nn\\
&+ \bar \eta_2 (|\phi_1|^2 + |\phi_2|^2)
\Big[\big(\phi_1^{*}\phi_2 \exp(i\theta_3)
+\phi_2^{*}\phi_1 exp(-i\theta_3) \big) \Big] \nn\\
&+ \bar \eta_3 (|\phi_1|^2 - |\phi_2|^2)
\Big[\big(\phi_1^{*}\phi_2 \exp(i\theta_4) \nn\\
&+\phi_2^{*}\phi_1 \exp(-i\theta_4) \big) \Big].
\label{jpot1}
\eea
Clearly the Josephson interaction breaks the
$U(1)\times U(1)$ symmetry of the potential (\ref{scpot2})
down to $U(1)$.
In general it is not easy to accommodate this type of
generalized Josephson interaction. But for simpler
Josephson interactions we can
accommodate them within the framework of the potential
(\ref{scpot2}). To understand how,
notice that the potential (\ref{scpot2}) already has
a Josephson interaction in the sense that it allows
an interband transition between $\phi_1$ and $\phi_2$
when $\lambda_{12}$ is not zero. This implies that
the above Josephson interaction could be included
in the $\lambda_{12}$ interaction.

\begin{figure*}[t]
\centering\subfigure[~~$|\phi_1|^2$]
{\includegraphics[scale=0.3]{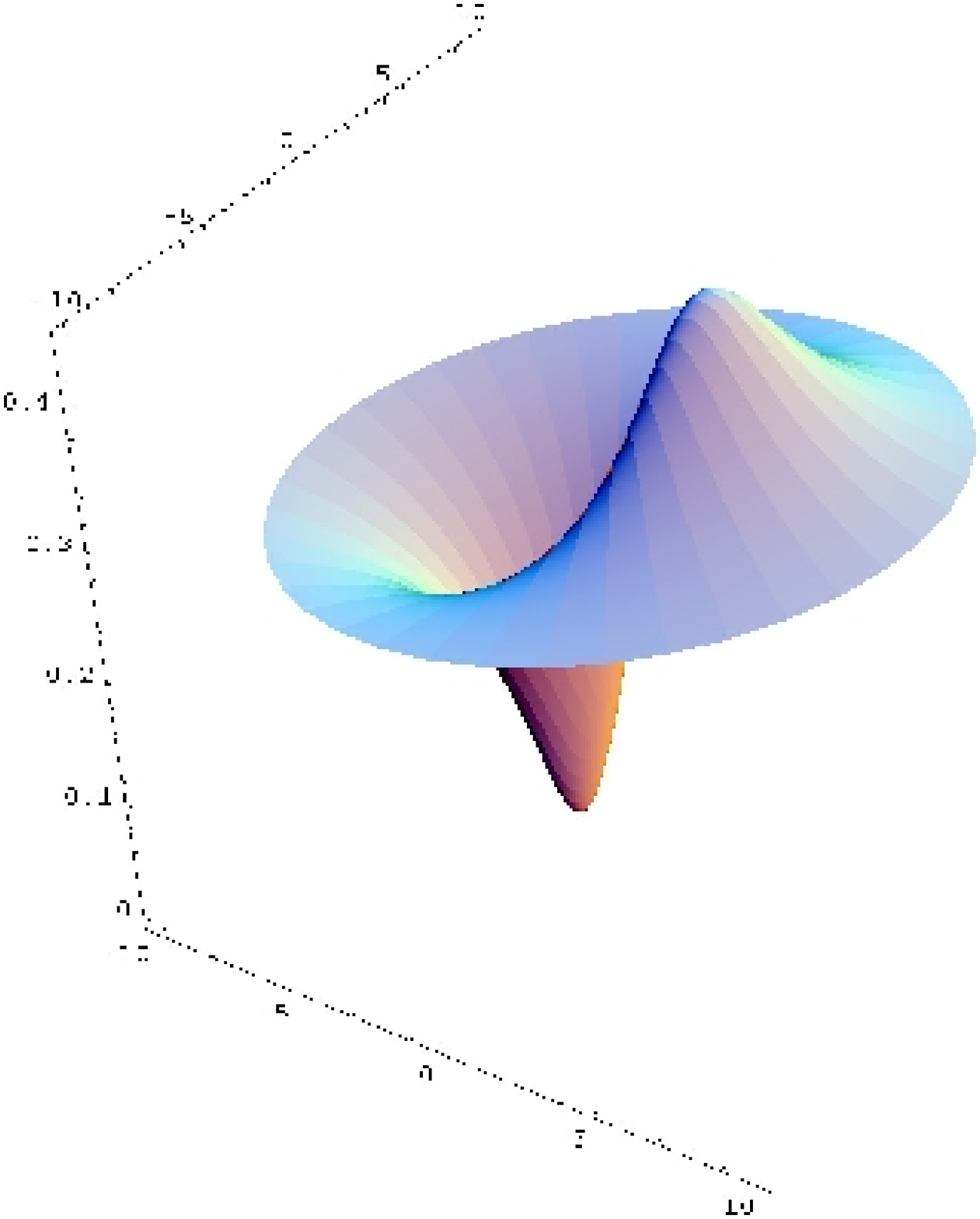}}\hfil
\subfigure[~~$|\phi_2|^2$]
{\includegraphics[scale=0.3]{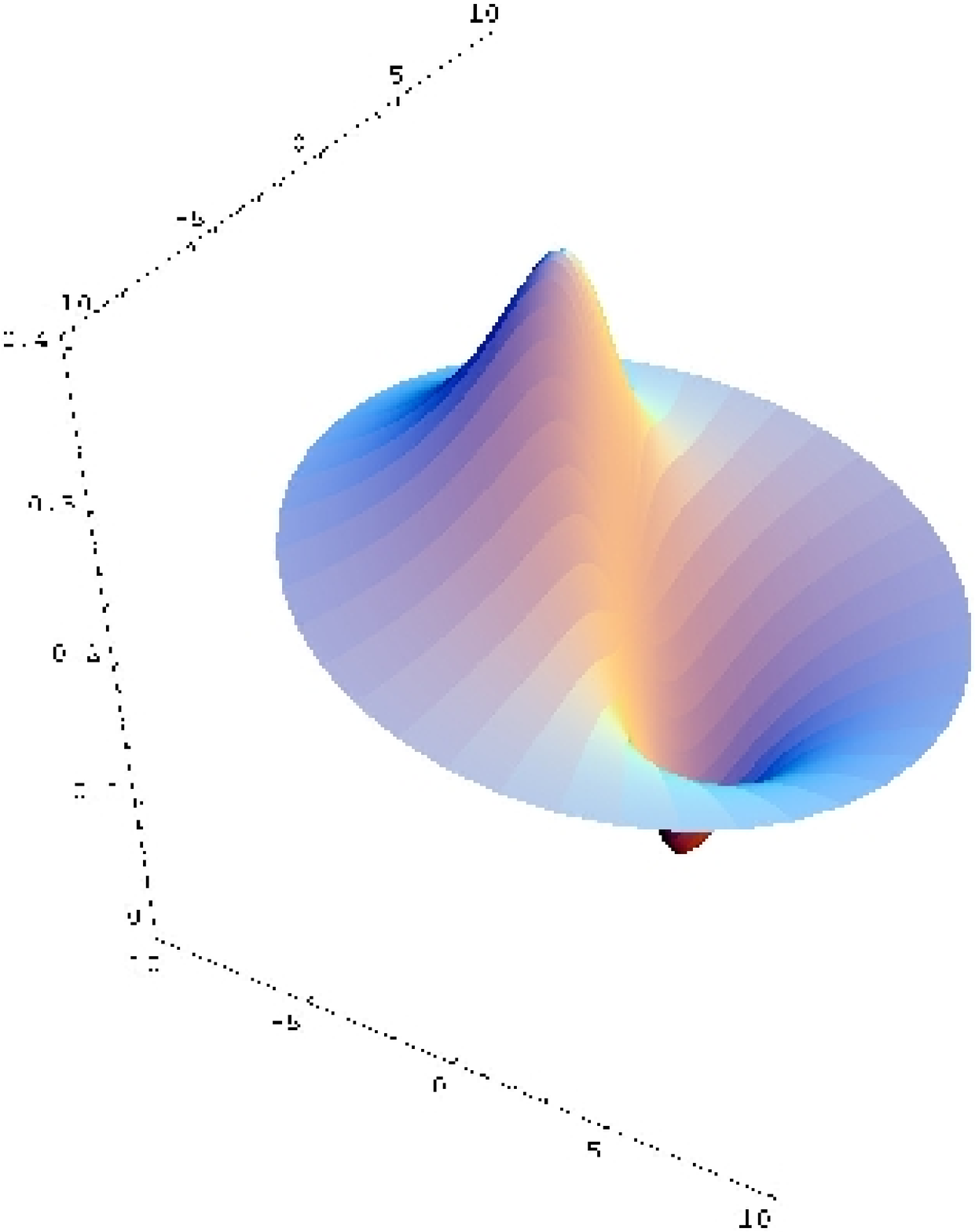}}
\caption{The density profile of $|\phi_1|^2$ and $|\phi_2|^2$ of
the N-type magnetic vortex in the presence of Josephson interaction.
Here we have put $\bar \rho=1$, $\gamma=0.05$, $\eta = 0.25$, and
$\lambda/g^2=2$.}
\label{jphi}
\end{figure*}

To exploit this point we generalize
the $U(1)\times U(1)$ symmetric potential (\ref{scpot2}) to
\bea
&V \rightarrow \bar V =V + V_1, \nn\\
&V_1 = \eta \Big[ \big(\phi_1^{*}\phi_2 \exp(i\theta)
+\phi_2^{*}\phi_1 \exp(-i\theta) \big) \Big] \nn\\
&+\eta_1 \Big[ \big(\phi_1^{*}\phi_2)^2 \exp(2i\theta)
+(\phi_2^{*}\phi_1)^2 \exp(-2i\theta) \big) \Big] \nn\\
&+ \eta_2 (|\phi_1|^2 + |\phi_2|^2)
\Big[\big(\phi_1^{*}\phi_2 \exp(i\theta)
+\phi_2^{*}\phi_1 \exp(-i\theta) \big) \Big] \nn\\
&+\eta_3 (|\phi_1|^2 - |\phi_2|^2)
\Big[\big(\phi_1^{*}\phi_2 \exp(i\theta) \nn\\
&+\phi_2^{*}\phi_1 \exp(-i\theta) \big) \Big],
\label{jpot2}
\eea
and introduce a new doublet $\psi$ with
an $SU(2)$ transformation of $\phi$,
\bea
&\psi = {\cal M} \phi, \nn\\
&{\cal M} = \Bigg( \matrix {\cos \dfrac a2 \exp(ib)
~~~-\sin \dfrac a2 \exp(ic) \cr
\sin \dfrac a2 \exp(-ic)~~\cos \dfrac a2 \exp(-ib) } \Bigg), \nn\\
&\tan a= \dfrac{\eta}\gamma.
\label{jt}
\eea
Now, we can show that when
\bea
&\theta=-b+c, \nn\\
&\eta_1= \dfrac{\eta^2}{2\gamma^2-\eta^2} \beta,
~~~~\eta_2 = -\dfrac{\eta}{2\gamma} \alpha, \nn\\
&\eta_3= \dfrac{2\gamma \eta}{2\gamma^2-\eta^2} \beta,
\label{jcon}
\eea
the potential (\ref{jpot2}) can be written as
\bea
&\bar V=\dfrac{\lambda_{11}'}{2}|\psi_1|^4+\lambda_{12}'|\psi_1|^2
|\psi_2|^2+\dfrac{\lambda_{22}'}{2}|\psi_2|^4 \nn\\
&-\mu_1'|\psi_1|^2-\mu_2'|\psi_2|^2 \nn\\
&=\dfrac {\lambda'} 2(|\psi _1|^2+|\psi _2|^2 -\dfrac{\mu'}{\lambda'})^2
+\dfrac {\alpha'} 2(|\psi _1|^4-|\psi _2|^4) \nn\\
&+\dfrac {\beta'}2 (|\psi_1|^2-|\psi _2|^2)^2
-\gamma' (|\psi _1|^2-|\psi _2|^2) \nn\\
&- \dfrac{\mu'}{2\lambda'},
\label{jpot3}
\eea
where
\bea
&\lambda' = \lambda - \eta_1, \nonumber \\
&\alpha'= \dfrac{\sqrt{\gamma^2+\eta^2}}\gamma \alpha,
~~~\beta' = \dfrac{2(\gamma^2+\eta^2)}{2\gamma^2-\eta^2} \beta,  \nn\\
&\mu' = \mu,~~~~~\gamma' = \sqrt{\gamma^2+\eta^2}.
\eea
So in terms of $\psi_1$ and $\psi_2$ the potential
(\ref{jpot2}) becomes the potential (\ref{scpot5})
which has no Josephson interaction.
This means that, with (\ref{jcon}), we can formally
absorb the Josephson interaction to $\lambda_{12}$ interaction.
From this we concludes that the presence of the Josephson
interaction does not affect the existence of the topological
objects in two-gap superconductor.

This does not mean that the Josephson interaction does not affect
the topological solutions. On the contrary, it does change
the shape of the solutions drastically. This is because
under the transformation (\ref{jt}) the profile of $\phi_1$
and $\phi_2$ change drastically.
To demonstrate this we let
$\alpha=\beta=0$ for simplicity, and adopt the potential
which has the following Josephson interaction
\bea
&V= \dfrac{\lambda}2 (|\phi_1|^2 + |\phi_2|^2-\dfrac{\mu}\lambda)^2
- \gamma (|\phi_1|^2 - |\phi_2|^2) \nn\\
&+\eta (\phi_1^{*}\phi_2 \exp(i\theta)
+\phi_1\phi_2^{*} \exp(-i\theta)).
\label{jpot}
\eea
Notice that, in terms of $\psi$, the potential is written as
\bea
&V= \dfrac{\lambda}2 (|\psi_1|^2 + |\psi_2|^2-\dfrac{\mu}\lambda)^2 \nn\\
&- \sqrt{\gamma^2+\eta^2}(|\phi_1|^2 - |\phi_2|^2),
\eea
so that it clearly has two types of straight vortex solution
of the following form
\bea
&\psi =\dfrac {\rho}{\sqrt{2}}
\Bigg( \matrix {\cos \dfrac f2 \exp(-in\varphi) \cr
\sin \dfrac f2 } \Bigg),
~~~\rho= \rho(\varrho),  \nn\\
&A_\mu=\dfrac ng A(\varrho) \pro_\mu \varphi.
\label{jvans1}
\eea
Now, in terms of $\phi$, the solution acquires the form
\bea
&\phi =\dfrac {\rho}{\sqrt{2}} \Big( \matrix {\xi_1 \cr
\xi_2} \Big), \nn\\
&\xi_1=\cos \dfrac f2 \cos \dfrac a2 \exp(-in\varphi-ib) \nn\\
&+ \sin \dfrac f2 \sin \dfrac a2 \exp(i\theta+ib), \nn\\
&\xi_2=-\cos \dfrac f2 \sin \dfrac a2 \exp(-in\varphi-i\theta-ib) \nn\\
&+\sin \dfrac f2 \cos \dfrac a2 \exp(ib).
\label{jvans2}
\eea
So, in the N-type vortex both $\phi_1$ and $\phi_2$ have a non-vanishing
concentration of at the core in the presence of the Josephson
interaction. The density profile of $\phi_1$ and $\phi_2$ of the
N-type $2\pi/g$-flux vortex with $b=0$ and $\theta=0$
is plotted in Fig.~\ref{jphi}.
Obviously the solution is not axially symmetric. More importantly
the vortex appears as a ``bound state" of two vortices
made of $\phi_1$ and $\phi_2$. This confirms that the
Josephson interaction does not prevent the the existence of
two types of magnetic vortex,
but changes the profile of the solutions drastically.
We notice that a similar vortex has been discussed in
two-component Bose-Einstein condensate \cite{ueda}.

Furthermore we can construct a helical vortex by
twisting the above vortex and making it periodic in $z$-coordinate.
In this case the helical vortex becomes a ``braided" magnetic vortex
made of two vortices of $\phi_1$ and $\phi_2$.
To see this we let
\begin{eqnarray}
&\phi= {\cal M}^{-1} \psi, \nn\\
&\psi =\dfrac {\rho}{\sqrt{2}} \left(
\begin{array}{l}
{\cos \dfrac f2 \exp{(-in\varphi) }} \\
{\sin \dfrac f2 \exp{(imkz)}}
\end{array}  \right), \nn\\
&A_\mu=\dfrac ng A_1(\varrho) \pro_\mu \varphi
+\dfrac{mk}g A_2(\varrho) \pro_\mu z,
\label{hjvans}
\end{eqnarray}
and again obtain two types of vortex. In this case the solution has
the following particle densities for $\phi_1$ and
$\phi_2$,
\bea
&|\phi _1|^2 =\dfrac{\rho ^2}4\Big( 1
+\dfrac{\gamma}{\sqrt{\gamma^2+\eta^2}} \cos f \nn\\
&+\dfrac{\eta}{\sqrt{\gamma^2+\eta^2}} \sin f
\cos (n\varphi+mkz+\theta+2b)  \Big), \nn \\
&|\phi _2|^2 =\dfrac{\rho ^2}4\Big( 1
-\dfrac{\gamma}{\sqrt{\gamma^2+\eta^2}} \cos f \nn\\
&-\dfrac{\eta}{\sqrt{\gamma^2+\eta^2}} \sin f
\cos (n \varphi+mkz +\theta+2b)  \Big).
\label{hjvp}
\eea
Clearly this shows that in the presence of the Josephson
interaction the helical magnetic vortex
becomes a braided vortex in which $\phi_1$-flux and $\phi_2$-flux
are braided together.

Now, it goes without saying that we can make a braided knot,
a twisted vortex ring, with the braided magnetic vortex.
This tells that the Josephson interaction makes
the topological objects in two-gap superconductor
more interesting.

\section{Non-Abelian Superconductor}

So far we have discussed an Abelian gauge theory of
two-gap superconductor. But our analysis implies that
the doublet ($\phi_1,\phi_2$) can be treated as an $SU(2)$
doublet. Indeed, when $\alpha=\beta=\gamma=0$,
the Lagrangian (\ref{sclag}) has
an exact $SU(2)$ symmetry. Even when there is no $SU(2)$
symmetry, one may still regard that the theory has an approximate
$SU(2)$ symmetry which is broken by the $\alpha$, $\beta$, and
$\gamma$ terms.
In this sense one may conclude that the Abelian two-gap superconductor
has a (broken) $SU(2)$ symmetry.
On the other hand the Ginzburg-Landau Lagrangian (\ref{sclag})
is still based on the Abelian electromagnetic interaction.
This leads us to wonder whether one can have
a genuine non-Abelian superconductor
in which the superconductivity is described by
a non-Abelian dynamics.

To discuss this issue, notice that in the above two-gap
superconductor the two condensates
$\phi_1$ and $\phi_2$ carry the same charge,
because the doublet is coupled to the Abelian
electromagnetic field. Now we show that
when the two condensates are made of opposite charges
(made of one electron-electron pair condensate and one hole-hole
pair condensate) the two-gap superconductor can be described by
a genuine non-Abelian $SU(2)$ gauge
theory. Moreover, we show that this type of non-Abelian superconductor
also allows a non-Abrikosov magnetic vortex and topological
knot identical to
what we have discussed in this paper.

To construct a theory of non-Abelian superconductivity which
is based on a genuine non-Abelian gauge theory, we need to understand
the mathematical structure of the non-Abelian gauge theory.
In non-Abelian gauge theory one can always decompose the
gauge potential into the restricted potential $\hat A_\mu$
and the valence potential $\X_\mu$.
Consider the $SU(2)$ gauge theory and let $\hn$ be
a gauge covariant unit triplet
which selects the charge direction of $SU(2)$. In this case
we have the following decomposition \cite{cho80,cho81},
\bea
& \vec{A}_\mu =A_\mu \n -
\oneg \n\times\pro_\mu\n+\X_\mu\nonumber
         = \hat A_\mu + \X_\mu, \nn\\
&  (A_\mu = \n\cdot \vec A_\mu,~ \n^2 =1,~
\hat{n}\cdot\vec{X}_\mu=0),
\label{cdecom}
\eea
where $ A_\mu$ is the
``electric'' potential. Notice that the restricted potential
is precisely the potential which leaves $\n$
invariant under the parallel transport,
\bea
\D_\mu \n = \pro_\mu \n
+ g {\hat A}_\mu \times \n = 0.
\eea
Under the infinitesimal
gauge transformation \bea \delta \n = - \vec \alpha \times \n
\,,\,\,\,\, \delta \A_\mu = \oneg  D_\mu \vec \alpha, \eea one has
\bea &&\delta A_\mu = \oneg \n \cdot \pro_\mu \valpha,\,\,\,\
\delta \hat A_\mu = \oneg \D_\mu \valpha  ,  \nn \\
&&\hspace{1.2cm}\delta \X_\mu = - \valpha \times \X_\mu  .
\eea
This tells three things. First, $\hat A_\mu$ by itself describes an $SU(2)$
connection which enjoys the full $SU(2)$ gauge degrees of
freedom. Secondly, the valence potential $\vec X_\mu$ forms a
gauge covariant vector field under the gauge transformation.
Most importantly, this tells that the decomposition is
gauge-independent. Once the gauge covariant topological field
$\hat n$ is given, the decomposition follows automatically
independent of the choice of a gauge \cite{cho80,cho81}.

The importance of the decomposition (\ref{cdecom}) for our
purpose is that one can construct a non-Abelian gauge theory,
a restricted gauge theory which has a full non-Abelian gauge
degrees of freedom, with the restricted potential
$\hat A_\mu$ alone \cite{cho80,cho81}. This is because
the valence potential $\vec X_\mu$ can be treated as
a gauge covariant source, so that one can exclude it from
the theory without compromising the gauge invariance.
Indeed it is this restricted gauge theory which describes
the non-Abelian gauge theory of
superconductivity \cite{prb05}.

Remarkably the restricted potential $\hat{A}_\mu$ retains
all the essential topological characteristics of the original
non-Abelian potential.
In fact, $\hat{n}$ defines the $\pi_2(S^2)$ topology
which describes the non-Abelian monopoles
and the $\pi_3(S^3)$ topology which characterizes
the topologically distinct vacua \cite{cho79,cho80,cho81}.
Furthermore it has a dual structure,
\begin{eqnarray}
& \hat{F}_{\mu\nu} = (F_{\mu\nu}+ H_{\mu\nu})\hat{n}\mbox{,}\nonumber \\
& F_{\mu\nu} = \partial_\mu A_{\nu}-\partial_{\nu}A_\mu \mbox{,}\nonumber \\
& H_{\mu\nu} = -\dfrac{1}{g} \hat{n}\cdot(\partial_\mu
\hat{n}\times\partial_\nu\hat{n}) = \partial_\mu
C_\nu-\partial_\nu C_\mu,
\end{eqnarray}
where $C_\mu$ is the ``magnetic'' potential.
Notice that this is exactly the potential $C_\mu$
that we have introduced in (\ref{id}).
This is an indication that the Ginzburg-Landau
theory of two-gap superconductor is closely related to
the restricted $SU(2)$ gauge theory.

With these preliminaries we now demonstrate a
non-Abelian superconductivity and non-Abelian Meissner effect.
Consider a $SU(2)$ gauge theory
described by the Lagrangian in which a doublet $\Phi$ couples
to the restricted $SU(2)$ gauge potential,
\bea
&{\cal L} = -|\hat D_\mu \Phi|^2 - V(\Phi,\Phi^{\dagger})
-\dfrac{1}{4} {\hat F}_{\mu\nu}^2, \nn\\
&\hat D_\mu \Phi = ( \partial_\mu + \dfrac{g}{2i} \vec \sigma
\cdot \hat A_\mu ) \Phi.
\label{nasclag1}
\eea
The equation of
motion of the Lagrangian is given by
\bea
&{\hat D}^2\Phi
=\dfrac{d V}{d \Phi^{\dagger}} \Phi, \nn\\
&\hat D_\mu \hat F_{\mu \nu}
= g \Big[(\hat D_\nu \Phi)^{\dagger}\dfrac{\vec\sigma}{2i} \Phi
- \Phi ^{\dagger} \dfrac{ \vec\sigma}{2i}(\hat D_\nu \Phi) \Big].
\label{nasceq}
\eea
Let $\xi$ and $\eta$
be two doublets which form an orthonormal basis,
\bea
&\xi{\dagger}\xi =1,~~~~~\eta^{\dagger} \eta = 1,
~~~~~\xi^{\dagger}\eta = \eta^{\dagger} \xi = 0, \nn\\
&\xi^{\dagger} \vec \sigma\xi = \hat n,
~~~~~\eta^{\dagger} \vec \sigma\eta= -\hat n, \nn\\
&(\hn \cdot \vec \sigma) ~\xi = \xi,
~~~~~(\hn \cdot \vec \sigma) ~\eta = -\eta,
\label{odbasis}
\eea
and let
\bea
\Phi = \phi_+ \xi + \phi_- \eta, ~~~~~(\phi_+= \xi^{\dagger}\Phi,
~~~\phi_-= \eta^{\dagger} \Phi).
\label{phidecom}
\eea
With this we have the identity
\bea
&\Big[\partial_\mu - \dfrac{g}{2i} \big(C_\mu \hn
+ \dfrac{1}{g} \hn \times \partial_\mu \hn \big)
\cdot \vec \sigma \Big] \xi = 0, \nn\\
&\Big[\partial_\mu + \dfrac{g}{2i} \big(C_\mu \hn
- \dfrac{1}{g} \hn \times \partial_\mu \hn \big)
\cdot \vec \sigma \Big] \eta = 0,
\label{cid0}
\eea
and find
\bea
\hat D_\mu \Phi = (D_\mu \phi_+) \xi + (D_\mu \phi_-)  \eta,
\label{phidef}
\eea
where
\bea
&D_\mu \phi_+ = (\partial_\mu + \dfrac{g}{2i} {\cal A}_\mu) \phi_+,
~~~D_\mu \phi_- = (\partial_\mu - \dfrac{g}{2i} {\cal A}_\mu) \phi_-, \nn\\
&{\cal A}_\mu = A_\mu + C_\mu, \nn\\
&C_\mu = \dfrac{2i}{g} \xi^{\dagger}\partial_\mu \xi
= - \dfrac{2i}{g} \eta^{\dagger}\partial_\mu \eta. \nn
\eea
From this we can express (\ref{nasclag1}) as
\bea
&{\cal L} = - |D_\mu \phi_+|^2 - |D_\mu \phi_-|^2
- V(\phi_+,\phi_-)  \nn\\
&- \dfrac{\lambda}{2} (\phi_+^{\dagger}\phi_+ + \phi_-^{\dagger}\phi_-)^2
-\dfrac{1}{4} {\cal F}_{\mu\nu}^2,
\label{nasclag2}
\eea
where
\bea
{\cal F}_{\mu\nu} = \partial_\mu {\cal A}_\nu
- \partial_\nu {\cal A}_\mu. \nn
\eea
This tells that the restricted $SU(2)$ gauge theory
(\ref{nasclag1}) is reduced to
an Abelian gauge theory coupled to oppositely charged
scalar fields $\phi_+$ and $\phi_-$. We emphasize that
this Abelianization is achieved without any gauge fixing.

The Abelianization assures that the non-Abelian Ginzburg-Landau
theory is not different from the Abelian Ginzburg-Landau theory.
Indeed with
\begin{equation}
\chi = \left(\begin{array}{rr}
\phi_+\\
\phi_-^*
\end{array}\right),
\end{equation}
we can express the Lagrangian (\ref{nasclag2}) as
\bea
&{\cal L} = - |D_\mu \chi|^2 + \mu^2 \chi^{\dagger}\chi
- \dfrac{\lambda}{2} (\chi^{\dagger} \chi)^2
- \dfrac{1}{4} {\cal F}_{\mu \nu}^2, \nn\\
&D_\mu \chi = (\partial_\mu + ig {\cal A}_\mu) \chi.
\label{nalag}
\eea
This is formally identical to the Lagrangian (\ref{sclag}) of
the Abelian two-gap superconductor discussed in Section III.
The only difference is
that here $\phi$ and $A_\mu$ are replaced by $\chi$ and ${\cal A}_\mu$.
This establishes that, with the proper redefinition of field
variables (\ref{phidecom}) and (\ref{phidef}), our non-Abelian
restricted gauge theory (\ref{nasclag1}) can in fact be made
identical to the Abelian gauge theory of two-gap superconductor. This
proves the existence of non-Abelian superconductors
made of the doublet consisting of oppositely charged
condensates \cite{prb05}. As importantly our analysis tells that
the two-gap Abelian superconductor has a hidden
non-Abelian gauge symmetry, because it can be transformed to
the non-Abelian restricted gauge theory. This implies that
the underlying dynamics of the two-gap superconductor
is indeed the non-Abelian gauge symmetry.
In the non-Abelian formalism it is explicit.
But in the Abelian formalism it is hidden,
where the full non-Abelian gauge symmetry only
becomes transparent when one embeds the doublet
$(\phi_1,\phi_2)$ properly into the non-Abelian symmetry.

Once the equivalence of two Lagrangians (\ref{sclag})
and (\ref{nasclag1}) is established, it must be evident
that the non-Abelian gauge theory of two-gap superconductor also
admits a non-Abrikosov vortex and magnetic knot.
This confirms the existence of a non-Abelian
Meissner effect and non-Abelian superconductivity.
All the above results of Abelian superconductor become
equally valid here.

\section{Discussion}

In this paper we have shown that the two-gap
superconductor can admit non-Abrikosov vortex
and topological knot. There are two types of non-Abrikosov vortex,
D-type and N-type. The D-type has no concentration of
the condensate at the core, but the N-type has a non-trivial profile of
the condensate at the core. In terms of topology
there are two, the non-Abelian topology $\pi_2(S^2)$
defined by $\hn$ and the Abelian topology
$\pi_1(S^1)$ defined by the invariant subgroup of $\hn$.
And both D-type and N-type vortices exist within the same
topological sector. In particular, we have infinitely many
D-type vortices which have the same topology.
The magnetic flux of the vortices can be integral or fractional.
The N-type vortex can have a $2\pi n/g$-flux or a fraction of this flux,
but the D-type vortex
has $2\pi k/g$ more flux than the N-type. And we have shown that
these non-Abrikosov vortices can be twisted
to form a helical vortex which is periodic in $z$-coordinate.

Perhaps a most interesting topological object in two-gap
superconductor is the magnetic knot, a twisted magnetic vortex
ring made of helical vortex. Our analysis suggests that
we have two types of knot, the D-type and the N-type.
They are made of two magnetic fluxes linked together,
one flux along the knot axis
and one flux along
the knot tube. And the linking number of two fluxes
provides the knot topology $\pi_3(S^2)$, which is described by
the Chern-Simon index of the electromagnetic potential.
The knot is stable dynamically as well as topologically.
The topological stability follows from the fact that
two flux rings linked together can not be separated by
any continuous deformation of the field configuration.
The dynamical stability follows from the fact that
the flux trapped inside of the knot ring can not be squeezed
out, which means that it provides
a repulsive force against the collapse of the knot.
Another way to understand this dynamical stability
is to notice that the supercurrent
along the knot generates a net
angular momentum around the knot axis.
And this provides the centrifugal
repulsive force preventing the knot to collapse.
This makes the knot dynamically stable.

The Josephson interaction makes these topological
objects more interesting. The straight vortex becomes
a bound state of two magnetic vortices made of two condensates
$\phi_1$ and $\phi_2$, and the helical vortex becomes
a braided magnetic vortex of two condensates.
Moreover the knot acquires the form of a braided magnetic
vortex ring. And we have two of them.

It must be emphasized, however, that the actual magnetic flux
of vortex and knot is determined by the two-gap superconductor
at hand because it is fixed by the parameters of
the potential which characterizes
the superconductor. Independent of this all two-gap superconductors
have two types of vortex and knot.
On the other hand one must keep the followings in mind.
First, compared with the N-type vortex the D-type vortex
has more energy in general because the D-type carries
more flux. This opens the possibility that,
within the same topological sector,
the D-type vortices could decay to the N-type vortices.
Secondly, the energy (per unit length) of
the fractional flux vortex and knot
is logarithmically divergent, so that they can exist only
when there is a cutoff parameter which makes the energy finite.
This tells that the N-type $2\pi/g$ vortex forms the true
finite energy ground state vortex of two-gap
superconductor.

Another important lesson from our analysis is that
the non-Abelian dynamics
could play a crucial role in condensed matter physics.
Indeed we have shown that we can actually
construct an $SU(2)$ gauge theory of superconductivity
which is mathematically equivalent to
the Abelian gauge theory of two-gap superconductor.
This means that, implicitly or explicitly, the underlying dynamics of
multi-gap superconductor can ultimately be related to
a non-Abelian gauge theory.

In this paper we have studied the topological objects in
two-gap superconductor. But from our discussion it must
be clear that similar topological objects should also exist
in multi-gap superconductor in general. This is because
the multi-gap superconductor is described by a multi-component
condensate, which naturally accommodates the non-trivial
non-Abelian topology.

Clearly the above theory of two-gap superconductor
is closely related to the Gross-Pitaevskii
theory of two-component Bose-Einstein
condensate (BEC), which tells that similar topological objects
can also exist in two-component BEC \cite{ijpap,ruo,pra05}.
This is because in the absence of the
electromagnetic interaction the above Ginzburg-Landau
Lagrangian reduces to the Gross-Pitaevskii Lagrangian
of two-component BEC. But there is an important difference.
In two-component BEC only the N-type vortex and knot
exist, because it allows only the N-type boundary
condition \cite{ijpap,pra05}.
In this sense it is really remarkable that two-gap
superconductor allows two types of topological objects.

We close with the following remarks: \\
1. Recently similar non-Abelian vortices and knots
have been asserted to exist almost everywhere, in atomic physics
in two-component BEC \cite{ijpap,ruo,pra05},
in condensed matter physics in multi-gap
superconductors \cite{ijpap,baba1},
in nuclear physics in Skyrme theory \cite{prl01,plb04},
in high energy physics in QCD \cite{plb05}.
The major difference here is that our vortex and knot are made of a
real magnetic flux. We emphasize that at the center of these
topological objects lies the baby skyrmion and the Faddeev-Niemi
knot in Skyrme theory. In fact, one can show that
our magnetic vortex and knot (as well as those in
two-component BEC) are a straightforward
generalization of the baby skyrmion and the Faddeev-Niemi
knot \cite{prl01,ijpap,plb04}. This is because both the Ginzburg-Landau
theory of two-gap superconductor and the Skyrme theory
are described by a $CP^1$ field $\hn$ which obeys the same
non-linear dynamics. \\
2. From our analysis there should be no doubt that the non-Abrikosov
vortex and the magnetic knot must exist in two-gap
superconductor. If so, the challenge now is
to verify the existence of these topological
objects experimentally. Constructing the knot
might not be a simple task at present moment. But the construction
of the non-Abrikosov vortex could be rather straightforward (at
least in principle) \cite{exp}.
To identify the non-Abrikosov vortex, there are two points
one has to keep in mind. First, the magnetic flux of the
non-Abrikosov vortex need not be $2\pi/g$, and can be fractional
in general. More importantly, there are two types of vortex,
the D-type which has no concentration of the condensate
at the core and the N-type which has
a non-trivial concentration of the condensate at the core.
These are the crucial points which distinguish the non-Abrikosov
vortex from the Abrikosov vortex.
With this in mind, one should be able to construct and identify
the non-Abrikosov vortex in two-gap superconductor
without much difficulty. \\
3. The non-Abelian gauge theory of superconductivity
is not just an academic curiosity. There is an excellent
example of non-Abelian two-gap
superconductor, the liquid metallic
hydrogen (LMH) \cite{ash}. Under high pressure the LMH becomes
a superconducting state in low temperature,
due to the electron Cooper pairs. But
in a lower temperature the proton Cooper pairs can coexist with
the electron pairs. And obviously it has no
Josephson interaction and probably a weak or no $\lambda_{12}$
interaction. So the LMH becomes an excellent
candidate of non-Abelian two-gap superconductor.
This implies that the LMH can have all
the topological objects we have discussed in this paper.
In particular it must have two types of
vortex and knot.

In this paper we have discussed
the topological objects which we obtain with the ansatz (\ref{svans}),
(\ref{hvans}), (\ref{sckans}), or (\ref{jvans2}). But we emphasize that
there are other topological objects which can be obtained
with different ansatz. These objects and the physical
implications of these topological objects in two-gap
superconductor will be discussed
in an accompanying paper \cite{sc5}.

{\bf ACKNOWLEDGEMENT}

~~~The work is supported in part by the ABRL Program of
Korea Science and Engineering Foundation (Grant R02-2003-000-10043-0).

\end{document}